\documentclass[namedreferences]{solarphysics}

\usepackage{placeins}
\usepackage{longtable}
\usepackage{array}
\usepackage[optionalrh]{spr-sola-addons} 
\usepackage{graphicx}        
\usepackage{amssymb}        
\usepackage{color}           
\usepackage{breakurl}  
\usepackage{amsmath}

\chardef\us=`\_

\begin{document}

\begin{article}
\begin{opening}

\title{Role of Non-Ideal Dissipation with Heating--Cooling Misbalance on the Phase Shifts of Standing Slow Magnetohydrodynamic Waves}

\author[addressref={aff1}]{Abhinav Prasad}
\author[addressref=aff1,corref,email={asrivastava.app@itbhu.ac.in}]{A.K. Srivastava}
\author[addressref=aff2]{Tongjiang Wang}
\author[addressref={aff1}]{Kartika Sangal}

\address[id=aff1]{Department of Physics, Indian Institute of Technology (BHU), Varanasi-221005, UP, India.}
\address[id=aff2]{The Catholic University of American and NASA Goddard Space Flight Center, Code 671, Greenbelt, MD, 20771, USA.}

\runningauthor{A. Prasad et al.}
\runningtitle{Phase Shift of Standing Slow Waves}

\begin{abstract}
We analyse the phase shifts of standing, slow magnetohydrodynamic (MHD) waves in solar coronal loops using a linear MHD model taking into account the role of thermal conductivity, compressive viscosity, radiative losses, and heating--cooling misbalance. We estimate the phase shifts in time and space of density and temperature perturbations with respect to velocity perturbations and also calculate the phase difference between density and temperature perturbations. The overall significance of compressive viscosity is found to be negligible for most of the loops considered in the study. For loops with high background density and/or low background temperature, the role of radiative losses (with heating--cooling misbalance) is found to be more significant. Also the effect of heating--cooling misbalance with a temperature- and density-dependent heating function is found to be more significant in the case of longer loop lengths ($L=500$\, Mm). We derived a general expression for the polytropic index [$\gamma_{\rm eff}$] and found that under linear MHD the effect of compressive viscosity on polytropic index is negligible. The radiative losses with constant heating lead to a monotonic increase of $\gamma_{\rm eff}$ with increasing density whereas the consideration of an assumed heating function [$H(\rho,T) \propto \rho^{a}T^{b}$, where $a=-0.5$ and $b=-3$] makes the $\gamma_{\rm eff}$ peak at a certain loop density. We also explored the role of different heating functions by varying the free parameters $a$ and $b$ for a fixed loop of $\rho_0 = 10^{-11}$\, kg $\text{m}^{-3}$, $T_0 = 6.3$\, MK and loop length $L= 180$\, Mm. We find that the consideration of different heating functions [$H(\rho,T)$] leads to a significant variation in the phase difference between density and temperature perturbations; however, the polytropic index remains close to a value of 1.66.   
\end{abstract}
\end{opening}

\section{Introduction}
     \label{S-Introduction} 
{
{\it Solar Ultraviolet Measurements of Emitted Radiation} (SUMER) spectrograph onboard the {\it Solar and Heliospheric Observatory} (SOHO) observed the Doppler-shift oscillations in coronal loops for the first time \citep{2002ApJ...574L.101W,2003A&A...406.1105W,2011SSRv..158..397W}. These SUMER oscillations were interpreted as the fundamental mode of the slow magnetoaccoustic oscillations \citep{2002ApJ...580L..85O,2003A&A...402L..17W}. Several other observations such as the {\it Yohkoh}/{\it Soft X-Ray Telescope} (SXT) suggest that the impulsive deposition of heat at footpoints of the loop is an important factor associated with the trigger of loop oscillations. Many localized transient events such as microflares lead to impulsive heating at the footpoint which eventually creates perturbations of both velocity and density within the loop  \citep{2005A&A...435..753W}. An inevitable response to the impulsive heating may also lead to associated pressure pulse and flows. Apart from the impulsive heating being suggested as primary excitation mechanism of slow magnetoaccoustic waves \citep{2005A&A...435..753W, 2006ApJ...647.1452P,2007ApJ...659L.173T}, there are various other mechanisms that can trigger the slow magnetoaccoustic oscillations, such as the kink instability or pressure pulses \citep[e.g.][and references therein]{2005A&A...436..701S,2007ApJ...668L..83S,2008A&A...479..235H}. 

Several quasi-periodic oscillations detected in stellar flares show features similar to those observed in solar flares \citep{2005A&A...436.1041M,2013ApJ...778L..28S,2016ApJ...830..110C}.
Quasi-periodic pulsations (QPPs) in solar and stellar flares are believed to be related to MHD oscillations and/or oscillatory reconnections \citep[c.f.][]{2004A&A...414L..25N,2006ApJ...644L.149O,2009SSRv..149..119N,2021SSRv..217...66Z}. \citet{2018PCF...61...N} suggested that a kind of QPPs that show damped harmonic-type oscillations are most likely caused by standing slow MHD modes in hot flaring loops. Damped QPPs are generally detected in the soft X-ray (SXR) and extreme-ultraviolet (EUV) emissions in solar flares, while they are more often detected in the white light in stellar flares \citep[e.g.][]{2015MNRAS.450..956B,2016MNRAS.459.3659P,2016ApJ...830..110C}. 

The {\it Solar dynamics observatory} (SDO)/{\it Atmospheric Imaging Assembly} (AIA) has more recently detected longitudinal intensity oscillations in flare loops. The observed decay times and period of these oscillations match the SUMER oscillations, and they have been suggested to be either a reflecting propagating slow-mode wave \citep[c.f.][]{2013ApJ...779L...7K,2015ApJ...804....4K,2016ApJ...828...72M,2017A&A...600A..37N} or standing slow wave mode \citep[c.f.][]{2015ApJ...811L..13W}. A linear, uniform loop model shows a sinusoidal temporal behaviour in the fundamental mode, whereas a reflecting wave does not have such a property. The AIA observations suggest that the perturbations are of the decaying sine-function form rather than quasi-periodic pulsations, thereby favouring the interpretation that the waves are of the standing nature.

\citet{2003A&A...406.1105W} performed a statistical study on the numerous loop oscillations detected by SUMER, and they were able to establish the physical properties of such oscillations. They found that the intensity fluctuation lags the Doppler shift by a quarter-period and the phase speed calculated from the observed period and loop length matches the local sound speed. \citet{2008A&A...481..247T} performed a numerical simulation that suggested that coronal loops maintained at 1 \,--\, 2\, MK temperature can also support slow-mode oscillations. The EUV {\it Imaging Spectrometer} (EIS) onboard {\it Hinode} \citep{2008ApJ...681L..41M, 2008A&A...489L..49E,2010NewA...15....8S} confirmed this later through observations. The slow-mode waves are strongly influenced by the dissipative agents e.g. thermal conductivity, compressive viscosity, and radiative losses. Thus a detailed study of the phase shifts of intensity and temperature perturbations can provide a useful diagnostics into the local solar atmosphere. A lot of progress has been made in the field of coronal seismology over the past decade providing the knowledge of parameters such as transport coefficients and polytropic index that are very important in the hydrodynamic as well as MHD modelling of solar and space plasmas  \citep[e.g.][for a review]{2015ApJ...811L..13W,2018ApJ...860..107W,2019ApJ...886....2W,2021SSRv..217...34W}. Recently \citet{2015ApJ...811L..13W} reported the first SDO/AIA observation of fundamental slow-mode oscillations, and they found that the phase speed matched the local sound speed in the hot loop at 9 \,MK. They calculated the polytropic index [$\gamma_{\rm eff}$] and found it to be close to the classical value of adiabatic index, which suggested that the thermal conductivity in hot coronal loops is suppressed. \citet{2015ApJ...811L..13W} used the observations of slow-mode waves and suggested that the observed rapid damping of slow MHD waves might be explained if we consider an enhancement in the classical compressive viscosity coefficient by more than an order of magnitude.
\citet{2019ApJ...886....2W} further refined this method using the numerical parametric modeling considering both thermal conduction and compressive viscosity.

The radiative-cooling and coronal-heating processes are also important in the evolution of slow-mode oscillations. The specific mechanism of chromospheric and coronal heating is still a widely discussed and studied topic in the solar context, and many theoretical studies have considered heating functions that depend on several plasma parameters, such as temperature, density, and magnetic field in order to analyse their role on coronal oscillations \citep[e.g.][and references therein]{2019A&A...628A.133K,2020A&A...644A..33K,2020SSRv..216..140V,2021SSRv..217...73N,2021SoPh..296...20P,2021JGRA..12629097S,2021SSRv..217...34W}. \citet{2019A&A...628A.133K} have performed a parametric study of different heating functions and highlighted the importance of heating--cooling misbalance in explaining the damping of standing slow magnetoaccoustic waves. Later \citet{2021SoPh..296...20P} also considered a specific heating function and suggested that the phenomenon of heating--cooling misbalance may lead to a better account for the period and damping of SUMER oscillations. In the present work we analyse the phase relationship between the temperature, density, and velocity perturbations of standing slow MHD oscillations for a wide range of coronal loops. \citet{2021SoPh..296..105P} have analysed the phase shifts of the slow-mode waves in a similar manner as done by \citet{2009A&A...494..339O} for the case of propagating waves. \citet{2021SoPh..296..105P} have also studied the phase shifts of propagating slow-mode waves using a MHD model taking into account the major dissipative effects along with a density- and temperature-dependent coronal heating function as proposed by \citet{2019A&A...628A.133K}. In their work, Fourier solutions of damped, propagating wave form were used to derive a dispersion relation of fourth order in the complex wave number [$k$]. As a follow up of their work, we present a parallel study for standing slow-mode waves where we have derived a generalized dispersion relation of third order in $\omega$, which is of complex nature in the damped, standing-wave solutions, while the wave number $k$ is real \citep[c.f.][and references therein]{2003A&A...408..755D,2006SoPh..236..127P,2007SoPh..246..187S,2019A&A...628A.133K}. \citet{2021SoPh..296..105P} compared the results of their theoretical model with the observations of \citet{2011ApJ...727L..32V} and \citet{2018ApJ...868..149K} thus providing an estimate of the coronal conditions that can account for the inferred polytropic index. In  their work \citet{2021SoPh..296..105P} focused on the warm coronal loops with equilibrium temperature in the range of $T_0 = 1$ \,--\, $2$\, MK and density ranging from $10^{-13}$ \,--\, $10^{-12}$\, kg ${\text m}^{-3}$. In the present work we compare our results for standing waves with the first SDO/AIA observation of fundamental, standing, slow modes studied in the work of \citet{2015ApJ...811L..13W}. We include a large parametric range of coronal loops having temperatures from $3$ \,--\, $10$\, MK, density from $10^{-12}$ \,--\, $10^{-10}$\, kg ${\text m}^{-3}$ and loop lengths in the range of $50$ \,--\, $500$\, Mm. Our comprehensive theory incorporates the major dissipative mechanisms, and we also provide a theoretical expression for the polytropic index which is used to analyse the effects of individual damping mechanisms. \citet{2019PhPl...26h2113Z} first studied the effect of misbalance between heating and cooling processes on polytropic index; however, in the present work we follow a different definition compared to their analysis. Finally we also explore the different range of heating functions and how they affect the phase shifts of slow MHD oscillations. Section 2 focuses on the MHD model and also provides the detailed theoretical expressions for the phase shifts of standing slow-mode oscillations. In various sub-sections of Section 3 we analyse the role of individual dissipative effects on the phase shifts. Particularly, in Section 3.6 we provide a brief comparison between the wave properties of standing and propagating slow-mode waves. Finally we discuss the implications and future prospects of the present study and also discuss some of its important extensions that can be developed in future.
}

\section{Basic Magnetohydrodynamic Model}
{
We model the slow-mode waves using MHD equations, which are simplified into a one-dimensional form by projecting the vector equations onto the stiff magnetic-field lines. We consider the role of thermal conductivity, viscosity, radiative losses, and heating--cooling misbalance in our model, while the gravitational stratification is ignored throughout our analysis. The coronal-heating function is considered to be density- and temperature-dependent of the form $H(\rho, T) \propto \rho^a T^b$ \citep{2019A&A...628A.133K} where $a$ and $b$ are free parameters of the heating function. We consider a homogeneous background plasma with equilibrium density, temperature, and pressure given as $\rho_0$, $T_0$, and $p_0$ respectively. The coronal loop length is equal to $L$. Similar to the analysis of \citet{2003A&A...408..755D}, the MHD equations are made dimensionless using the background-plasma parameters and a length scale equal to $2L$. The corresponding time scale is thus given as $\tau = 2L/c_{\rm s0}$ where $c_{\rm s0} = \sqrt{\gamma p_0/\rho_0}$ is the sound speed. Further we linearize the MHD equations as written below. Note that $v_1$, $\rho_1$, $T_1$, and $p_1$ are the dimensionless first order perturbations.
\begin{align}
    \frac{\partial \rho_1}{\partial t} &= -\frac{\partial v_1}{\partial z},\\
    \frac{\partial v_1}{\partial t} &= -\frac{1}{\gamma}\frac{\partial p_1}{\partial z} + e\frac{\partial^2 v_1}{\partial z^2},\\
    \frac{\partial T_1}{\partial t} &= -(\gamma-1)\frac{\partial v_1}{\partial z} + \gamma d  \left(  \frac{\partial^2 T_1}{\partial z^2} \right) - \gamma r(\alpha T_1 + \rho_1) + \gamma r(bT_1 + a\rho_1),\\
    p_1&=  \rho_1 +  T_1,
\end{align}
here the dimensionless ratios are defined as in \citet{2003A&A...408..755D},
\begin{align}
    e &= \frac{2\eta_0 T_{0}^{5/2}}{3\rho_{0}L c_{\rm s0}}\,\,\, \text{(viscous ratio)},\\
    d &=\frac{(\gamma-1)\kappa_0T_{0}^{7/2}\rho_0c_{\rm s0}}{2\gamma^2 p_0^2L}\,\,\, \text{(thermal ratio)},\\
    r &= \frac{2(\gamma-1)L \rho_{0}^2}{\gamma p_0 c_{\rm s0}} \chi T_{0}^{\alpha}\,\,\, \text{(radiative  ratio)}.
\end{align}
The equilibrium parameters satisfy the ideal gas equation as below,
\begin{equation}
    \rho_{0} = \frac{mp_{0}}{{\text k}_{\rm B}T_{0}}.
\end{equation}
where $m$ is the mean particle mass and ${\text k}_{\rm B}$ is the Boltzmann constant.\newline
Further we have 
\begin{equation*}
    \kappa_0 = 9 \times 10^{-12}\,\, {\text W}\, {\text m}^{-1}\,{\text K}^{-1}\,\,\,;\,\,\,\eta_0 = 10^{-17}\,\, \text{kg} \,\,{\text m}^{-1}\,{\text s}^{-1}.
\end{equation*}
We have considered two different approximate models of the radiative cooling-function (valid in coronal abundance) for our study \citep{2008ApJ...682.1351K,2014masu.book.....P},
\begin{align}
    \text{Model I:}\,\,\, \chi T^{\alpha} &= \frac{10^{-32}}{m^2} T^{-1/2} \,\,\,\,\, (10^6\,\, {\rm K} < T < 10^7\,\, {\rm K})\\
    \text{Model II: }\,\,\, \chi T^\alpha &=
    \begin{cases}
   \frac{ 3.46 \times 10^{-38}}{m^2} T^{1/3}\, ,\,\,\,\, 10^{6.55}\,\, {\rm K} < T \leq 10^{6.9}\,\, {\rm K}\\
    \frac{5.49 \times 10^{-29}}{m^2} T^{-1} \, ,\,\,\,\,\,10^{6.9} \,\,{\rm K} < T < 10^{7.63}\,\, {\rm K}
    \end{cases}
\end{align}
here the radiative cooling function(s) has been specified in SI units of ${\rm W} {\text m}^{3} \text{ kg}^{-2}$.
In Table 1 we have summarised the thermal, viscous, and radiative ratios for the entire range of loops considered in our study. These parameters are calculated as
\begin{align}
    e &= \frac{2\eta_0}{3L}\left( \frac{m}{\gamma {\text k}_{\rm B}} \right)^{1/2} \frac{T_0^2}{\rho_0} = 4.402 \times 10^{-20} \frac{T_0^2}{L\rho_0},\\
    d &= \frac{(\gamma-1)\kappa_0}{2L}\left( \frac{m}{\gamma {\text k}_{\rm B}} \right)^{3/2}\frac{T_0^2}{\rho_0} = 8.641 \times 10^{-19}\frac{T_0^2}{L\rho_0},\\
    r &= 2L(\gamma-1)\chi\left( \frac{m}{\gamma {\text k}_{\rm B}} \right)^{3/2} \frac{\rho_0}{T_0^2} = 3.816 \times 10^{15} \frac{L\rho_0}{T_0^2}\,\,\,\,\,\, \text{[Model I]},\\
   r&=
    \begin{cases}
       1.32 \times 10^{10} \frac{L\rho_0}{T_0^{7/6}} & \, ,\,\,\,\, 10^{6.55}\,\, {\rm K} < T_0 \leq 10^{6.9}\,\, {\rm K}\\
       2.09 \times 10^{19} \frac{L\rho_0}{T_0^{5/2}} & \, ,\,\,\,\, 10^{6.9}\,\, {\rm K} < T_0 \leq 10^{7.63}\,\, {\rm K}
    \end{cases}
    \,\,\,\,\,\text{[Model II]},
\end{align}
where $T_0$, $\rho_0$, and $L$ are in SI units. \newline
\begin{table*}
\begin{tabular}{p{1.5cm} p{3cm} p{1.35cm} p{1.35cm} p{1.2cm} p{1.2cm}}

\hline
 {\bf Loop length} & {\bf Loop Parameters} & {\bf Thermal Ratio [$d$]} & {\bf Viscous Ratio [$e$]} & \multicolumn{2}{c}{\bf Radiative ratio [$r$]} \\
 & & & & Model I & Model II\\
\hline
 50\, Mm & $T_0$ = $T_{00}$, $\rho_0 = \rho_{00}$  & 0.043& 0.0022 & 0.076 & 0.1009 \\
 & $T_0$ = 0.6 $T_{00}$, $\rho_0 = \rho_{00}$  & 0.015& 0.0007 & 0.212 & 0.183\\
 &  $T_0$ = 2 $T_{00}$, $\rho_0 = \rho_{00}$ & 0.172 & 0.0088 & 0.019 & 0.033\\
 &  $T_0$ = $T_{00}$, $\rho_0 = 0.1\rho_{00}$ & 0.432 & 0.022 & 0.0076 & 0.01\\
 &  $T_0$ = $T_{00}$, $\rho_0 = 10 \rho_{00}$ & 0.00432 & 0.0002 & 0.763 & 1.009 \\
\hline
 180\, Mm & $T_0$ = $T_{00}$, $\rho_0 = \rho_{00}$  & 0.012& 0.0006 & 0.27 & 0.363 \\
 & $T_0$ = 0.6 $T_{00}$, $\rho_0 = \rho_{00}$  & 0.0043& 0.00022 & 0.763 & 0.659 \\
 &  $T_0$ = 2 $T_{00}$, $\rho_0 = \rho_{00}$ & 0.048 & 0.0024 & 0.068 & 0.119\\
 &  $T_0$ = $T_{00}$, $\rho_0 = 0.1\rho_{00}$ & 0.12 & 0.0061 & 0.027 & 0.036 \\
 &  $T_0$ = $T_{00}$, $\rho_0 = 10\rho_{00}$ & 0.0012 & 0.00006 & 2.747 & 3.63\\
\hline
 300\, Mm & $T_0$ = $T_{00}$, $\rho_0 = \rho_{00}$  & 0.007& 0.00036 & 0.457 & 0.605 \\
 & $T_0$ = 0.6 $T_{00}$, $\rho_0 = \rho_{00}$  & 0.0025& 0.00013 & 1.27 & 1.09 \\
 &  $T_0$ = 2 $T_{00}$, $\rho_0 =  \rho_{00}$ & 0.0288 & 0.0014 & 0.114 & 0.198\\
 &  $T_0$ =  $T_{00}$, $\rho_0 = 0.1\rho_{00}$ & 0.072 & 0.0036 & 0.045 & 0.06 \\
 &  $T_0$ = $T_{00}$, $\rho_0 = 10\rho_{00}$ & 0.0007 & 0.00003 & 4.57 & 6.05 \\
\hline
 500\, Mm & $T_0$ = $T_{00}$, $\rho_0 = \rho_{00}$  & 0.0043& 0.00022 & 0.763 & 1.0098 \\
 & $T_0$ = 0.6 $T_{00}$, $\rho_0 = \rho_{00}$  & 0.0015& 0.00007 & 2.12 & 1.83\\
 &  $T_0$ = 2 $T_{00}$, $\rho_0 =  \rho_{00}$ & 0.017 & 0.0008 & 0.19 & 0.331\\
 &  $T_0$ =  $T_{00}$, $\rho_0 = 0.1\rho_{00}$ & 0.043 & 0.0022 & 0.076 & 0.1009 \\
 &  $T_0$ = $T_{00}$, $\rho_0 = 10 \rho_{00}$ & 0.00043 & 0.00002 & 7.63 & 10.098 \\
\hline
\end{tabular}
\caption{Summary of the dimensionless ratios for the entire range of loop lengths and parameters considered in the study. ($T_{00} = 5$\,MK and $\rho_{00} = 10^{-11}$\, kg \,${\text m}^{-3}$)}
\end{table*}
}

\subsection{Dispersion Relation} 
{
Considering the perturbations to be of the form
\begin{equation}
    f \propto {\text e}^{{\rm i}(kz-\omega t)}.
\end{equation}
We shall substitute the above Fourier solutions in the linearized MHD equations, further since we model the standing slow modes thus the solutions are considered to be damped with increasing time (complex $\omega$) while the wave number $k$ is real making the solutions of the standing-wave form \citep[c.f.][]{2003A&A...408..755D,2007SoPh..246..187S}. Finally we obtain a dispersion relation that is cubic in $\omega$ and quartic in $k$ \citep[c.f.][in detail]{2021SoPh..296..105P}. We shall solve it for a fixed wavenumber given as $k = 2n\pi$, where $n=1,2,3...$ correspond to fundamental and higher-order harmonics. The dispersion relation is given as,
\begin{equation}
    \omega^3 + A\omega^2 + B\omega + C =0,
\end{equation}
where 
\begin{align}
    A &= -{\rm i}(\gamma r (b - \alpha) - \gamma d k^2) + {\rm i}ek^2,\\
    B &= ek^2(\gamma r (b - \alpha) - \gamma d k^2) - k^2,\\
    C &= {\rm i}k^2( r (b - \alpha) - d k^2) - {\rm i}k^2 r (a-1).
\end{align}
It is mentioned here that \citet{2021SoPh..296..105P} derived the same dispersion relation; however, in their work it was considered for the case of propagating, slow-mode waves where the Fourier solutions had a fixed real $\omega$ and the dispersion relation was solved for complex wavenumber $k$. In the present analysis we build on their work and study the nature of standing slow-mode waves.  
Since $\omega$ is complex for standing modes, we write it as $\omega = \omega_{\rm r} + {\rm i}\omega_{\rm i} = \omega_{\rm m} {\text e}^{{\rm i}\phi}$ where $\omega_{\rm m} = \sqrt{\omega_{\rm r}^2 + \omega_{\rm i}^2}$ and $\phi = \tan^{-1}\left(\frac{\omega_{\rm i}}{\omega_{\rm r}}\right)$. On solving the dispersion relation we find two complex roots of the form $\pm \omega_{\rm r} + {\rm i}\omega_{\rm i}$ corresponding to the slow, standing wave modes and a third root with $\omega_{\rm r} \approx 0$ corresponding to the thermal mode. \newline
}
\subsection{Phase Relations} 
{
The velocity perturbations of standing slow MHD modes can be given as a sum of two opposite-propagating wave solutions,
\begin{equation}
    v_1 = \hat{v}_1{\text e}^{\omega_{\rm i} t}({\text e}^{{\rm i}(kz-\omega_{\rm r} t)} + {\text e}^{{\rm i}(-kz - \omega_{\rm r} t)}).
\end{equation}
The real part of above equation is thus
\begin{equation}
    v_1 = \hat{v}_1\cos(\omega_{\rm r} t)\cos(kz){\text e}^{\omega_{\rm i} t}.
\end{equation}
Substituting Equation 20 in the continuity equation we obtain
\begin{equation}
    \frac{\partial \rho_1}{\partial t} = {\rm i}k\hat{v}_1{\text e}^{\omega_{\rm i} t}({\text e}^{{\rm i}(-kz-\omega_{\rm r} t)} - {\text e}^{{\rm i}(kz - \omega_{\rm r} t)}).
\end{equation}
Further integrating we have
\begin{equation}
    \rho_1 = \frac{{\rm i}k(\omega_{\rm i} + {\rm i}\omega_{\rm r})\hat{v}_1}{\omega_{\rm m}^2}{\text e}^{\omega_{\rm i} t}({\text e}^{{\rm i}(-kz-\omega_{\rm r} t)} - {\text e}^{{\rm i}(kz - \omega_{\rm r} t)}).
\end{equation}
Taking the real part of above equation we finally obtain
\begin{equation}
    \rho_1 = \frac{2k\hat{v}_1}{\omega_{\rm m}}\cos\left(\omega_{\rm r} t + \phi-\frac{\pi}{2}\right)\cos\left(kz-\frac{\pi}{2}\right){\text e}^{\omega_{\rm i} t}.
\end{equation}
Comparing Equations 21 and 24 we find that the phase shift of $\rho_1$ with respect to $v_1$ in time due to non-ideal effects is given by $\phi$. Further there is also an extra phase difference of $\frac{\pi}{2}$ in both time and space for the standing slow-mode waves. In addition, it shows that the non-ideal dissipations do not affect the $\frac{\pi}{2}$ phase relationship between $v_1$ and $\rho_1$ in space for the standing wave. The phase shift in time of $\rho_1$ with respect to $v_1$ is given by

\begin{equation}
    \phi_{\rho} = \frac{\pi}{2}-\phi = \frac{\pi}{2}-\tan^{-1}\left( \frac{\omega_{\rm i}}{\omega_{\rm r}} \right).
\end{equation}

We have accordingly defined $\phi_\rho$ being the phase difference of the variation of density perturbations compared to the velocity perturbations. The phase relation between $v_1$ and $\rho_1$ for damped, standing, slow-mode waves was also derived using the similar method previously by \citet{2003A&A...402L..17W} and \citet{2006SoPh..236..127P}. $\phi_\rho$ will be regarded as the density phase shift hereafter.

From the momentum conservation equation we can express the temperature perturbations in terms of velocity and density perturbations,
\begin{equation}
     \frac{\partial T_1}{\partial z} = \gamma e\frac{\partial^2 v_1}{\partial z^2} - \gamma\frac{\partial v_1}{\partial t} - \frac{\partial \rho_1}{\partial z}.
\end{equation}
Thus, substituting Equations 20 and 23 into the above equation and integrating we get
\begin{equation}
    T_1 = \left(-{\rm i}\gamma ek - \frac{{\rm i}\gamma(\omega_{\rm i} - {\rm i}\omega_{\rm r})}{k} - \frac{{\rm i}k(\omega_{\rm i} + {\rm i}\omega_{\rm r})}{\omega_{\rm m}^2}\right)\hat{v}_1 ({\text e}^{{\rm i}(-kz-\omega_{\rm r} t)} - {\text e}^{{\rm i}(kz - \omega_{\rm r} t)}){\text e}^{\omega_{\rm i} t}.
\end{equation}
Finally taking the real part of above equation we get
\begin{equation}
    T_1 = 2R\cos\left(\omega_{\rm r} t - \Phi-\frac{\pi}{2}\right)\cos\left(kz-\frac{\pi}{2}\right){\text e}^{\omega_{\rm i} t},
\end{equation}
where
\begin{equation}
    R = \sqrt{((\alpha_1 - \beta)\cos\phi)^2 + (\beta_1 + (\alpha_1 + \beta)\sin\phi)^2},
\end{equation}
\begin{equation}
    \Phi = \tan^{-1}\left( \frac{\beta_1 + (\alpha_1 + \beta)\sin\phi}{(\alpha_1 - \beta)\cos\phi} \right),
\end{equation}
and 
\begin{equation}
    \alpha_1 = \frac{\gamma \omega_m }{k} \,\,\,,\,\,\,\, \beta = \frac{ k}{\omega_m}\,\,\,,\,\,\,\, \beta_1 = \gamma e k.
\end{equation}
Comparing Equations 21 and 28 we find that the phase shift in time of $T_1$ with respect to $v_1$ due to non-ideal effects is given by $\Phi$. The comparison also indicates that there is an additional $\frac{\pi}{2}$ phase difference in both time and space for the same reason as mentioned above. The phase shift in time of $T_1$ with respect to $v_1$ is thus similarly given as

\begin{equation}
    \phi_{\rm T} = \frac{\pi}{2}+\Phi.
\end{equation}

Hereafter we will mention $\phi_{\rm T}$ as the temperature phase shift. It is clear that in the ideal MHD case (i.e. $e=d=r=0$, so $\omega_{\rm i} = 0$) using Equation 30 we have $\phi=\Phi=0$, so $\rho_1$ and $T_1$ have a $\frac{\pi}{2}$ phase shift with $v_1$ in both time and space as it has been well known for a compressive standing wave. \citep[e.g.][]{Rayleigh_book,2002SoPh..209..265S}.}

\section{Analysis of Theoretical Results} 
{
In subsequent sections we study the role of different damping mechanisms on the phase shift of standing slow-mode waves in a step-wise manner. We study a wide range of loop densities (  $10^{-12}$ \,--\, $10^{-10}$\, kg \,${\text m}^{-3}$) and temperatures ( 3 \,--\, 10\, MK) along with a discrete range of loop lengths from shorter loops of $L=50$\, Mm to longer loops of $L=500$\, Mm. The individual effect of damping mechanisms on the polytropic index is studied in Section 3.5. In Section 3.6 we provide a brief comparison between the analysis of phase shift and polytropic index of propagating and standing slow-mode waves and summarize the role of heating--cooling misbalance on the phase shifts of slow modes. It is also mentioned here that all of the cases and coronal loops studied in the following subsections are those having stable standing slow-mode and thermal-mode solutions given by the dispersion relation, i.e. $\omega_{\rm i} < 0$.
}
  
\subsection{Role of Thermal Conductivity} 
{
 \begin{figure}   
   \centerline{\hspace*{0.03\textwidth}
               \includegraphics[width=0.56\textwidth,clip=]{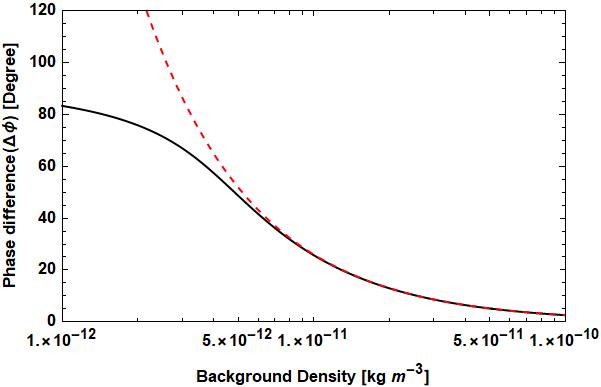}
               \hspace*{0.06\textwidth}
               \includegraphics[width=0.56\textwidth,clip=]{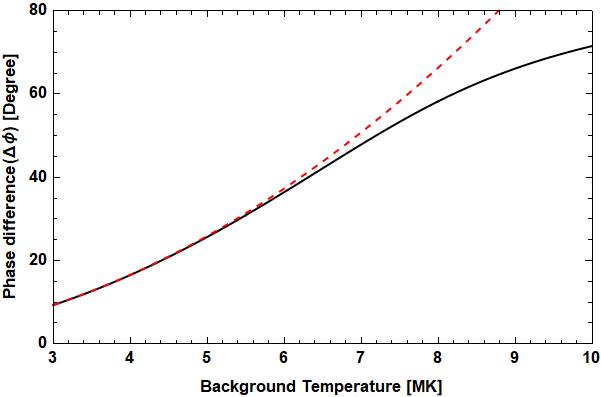}
              }
     \vspace{-0.35\textwidth}   
     \centerline{\Large \bf     
      \hspace{-0.17 \textwidth}  \color{black}{\footnotesize{(a)}}
      \hspace{0.58\textwidth}  \color{black}{\footnotesize{(b)}}
         \hfill}
     \vspace{0.31\textwidth}    
   \centerline{\hspace*{0.015\textwidth}
               \includegraphics[width=0.56\textwidth,clip=]{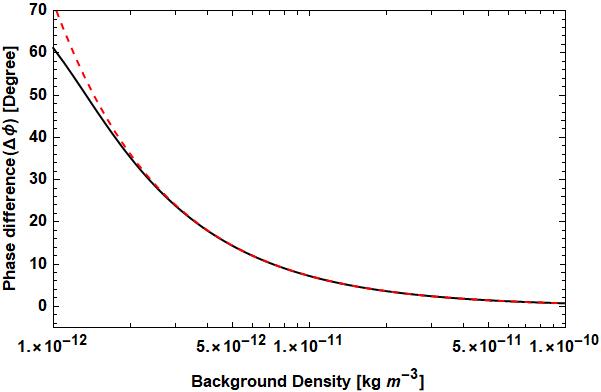}
               \hspace*{0.06\textwidth}
               \includegraphics[width=0.56\textwidth,clip=]{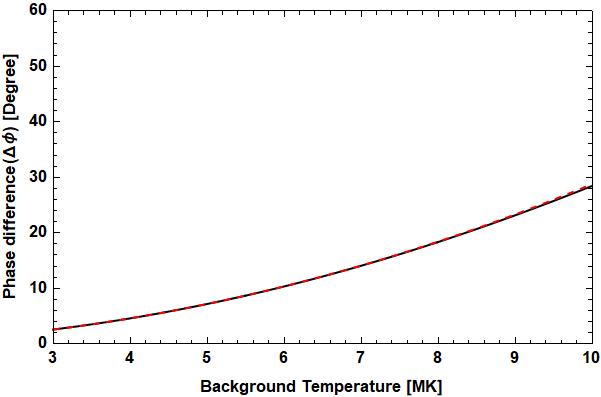}
              }
     \vspace{-0.35\textwidth}   
     \centerline{\Large \bf     
      \hspace{-0.17 \textwidth} \color{black}{\footnotesize{(c)}}
      \hspace{0.58\textwidth}  \color{black}{\footnotesize{(d)}}
         \hfill}
     \vspace{0.31\textwidth}    
\caption{Variations of phase difference of fundamental mode with respect to the background density and temperature for loop length of $L=50$\, Mm ({\it top panels}) and $L=180$\, Mm ({\it bottom panels}). For calculation of curves in the {\it left column}, $T_0=5$ MK is used, and for the {\it right column} $\rho_0=10^{-11}$\,kg\, ${\text m}^{-3}$ is used. The phase difference calculated using Equation 37 is represented by {\it black-solid} curves and the {\it red-dashed} curves represent the weak-damping approximations (Equation 44).} 
 \label{F-4panels}
 \end{figure}
In the present section we consider only thermal conductivity as a damping mechanism in the MHD equations. Therefore we simplify the dispersion relation (Equation 16) by only considering the effect of thermal conductivity ($e=r=0$)
\begin{equation}
    \omega^3 + A\omega^2 + B\omega + C =0,
\end{equation}
where 
\begin{align}
    A &= {\rm i}\gamma k^2 d,\\
    B &= -k^2,\\
    C &= -{\rm i}dk^4.
\end{align}
This expression is consistent with Equation 19 in \citet{2003A&A...408..755D}. Note that the difference in signs of the coefficients is due to the definition of ${\text e}^{{\rm i}(\omega t-kz)}$ in \citet{2003A&A...408..755D}.
Using Equations 25 and 32 we write the phase difference between temperature and density perturbation as 

\begin{equation}
    \Delta \phi = \phi_{\rho} - \phi_{\rm T} = -(\Phi+\phi).
\end{equation}
We will call $\Delta{\phi}$ simply the phase difference from hereafter.
\newline
Further considering the density and temperature perturbations as
\begin{equation}
    \rho_1 = \hat{\rho}_1({\text e}^{{\rm i}(kz-\omega t)} + {\text e}^{{\rm i}(-kz-\omega t)}),
\end{equation}
\begin{equation}
    T_1 = \hat{T}_1 ({\text e}^{{\rm i}(kz-\omega t - \Delta \phi)}+{\text e}^{{\rm i}(-kz-\omega t - \Delta \phi)} ).
\end{equation}
The density and temperature perturbations are substituted in the energy conservation equation, which simplifies as below under the effect of thermal conductivity only:
\begin{equation}
     \frac{\partial T_1}{\partial t} = (\gamma-1)\frac{\partial \rho_1}{\partial t} + \gamma d  \left(  \frac{\partial^2 T_1}{\partial z^2} \right).
\end{equation}
We thus have
\begin{equation}
    {\text e}^{-{\rm i}\Delta \phi}\left(\frac{{\rm i}\gamma d k^2}{\omega} + 1\right)\hat{T}_1 =  (\gamma-1)\hat{\rho}_1,
\end{equation}
\begin{equation}
    \left( \frac{{\rm i}\gamma d k^2 }{\omega_{\rm r}^2 + \omega_{\rm i}^2}(\omega_{\rm r} - {\rm i}\omega_{\rm i}) + 1 \right)(\cos \Delta \phi - {\rm i}\sin \Delta \phi)\hat{T}_1 = (\gamma-1)\hat{\rho}_1.
\end{equation}
The imaginary part of Equation 42 is thus written as
\begin{equation}
    \tan \Delta \phi = \frac{\gamma d \omega_{\rm r} k^2/(\omega_{\rm r}^2 + \omega_{\rm i}^2)}{1+ (\gamma d k^2 \omega_{\rm i})/(\omega_{\rm r}^2 + \omega_{\rm i}^2)}.
\end{equation}
The above equation is equivalent to the similar expression derived by \citet{2018ApJ...860..107W}. Note that in their analysis the Fourier solutions were taken of the form ${\text e}^{{\rm i}(\omega t - kz)}$ and the thermal ratio $d$ was defined using a timescale different from the present analysis. Under the weak-damping assumption $\omega_{\rm i}\approx 0$, $\omega_{\rm r}\approx k$, we get
\begin{equation}
\tan\Delta\phi=\gamma d k.
\end{equation}
Note that above expression is same as Equation 12 in \citet{2018ApJ...860..107W} for the fundamental mode $k=2\pi$. We numerically solve the dispersion relation (Equation 33) using the Wolfram {\it Mathematica} Environment from 2016 for the fundamental mode $k=2\pi$. Note that throughout our analysis we will be talking about the fundamental mode only.

Panels of Figure 1 show the phase difference between density and temperature perturbations obtained by substituting the numerical solution of Equation 33 for fundamental mode ($k=2\pi$) into Equation 37. The top panels are plotted for a loop length of $50$\, Mm while the bottom panels are for $L = 180$\, Mm. The left panels in each row are plotted with respect to the background density while considering a constant temperature of $T_0 = 5$\, MK, and similarly the corresponding right panels are with respect to background temperature at a constant density of $\rho_0 = 10^{-11}$\, kg\, ${\text m}^{-3}$. The calculations show that the numerical solution (black-solid curve) and the analytical solution given by Equation 43 (not shown) match completely. We compare the numerical solution (black-solid curve) with the analytical approximation (red-dashed curve) given by Equation 44. The panels show that the analytical approximation matches well with the numerical solutions for large loop lengths ($L=180$\, Mm). This is expected because the thermal ratio $d \ll 1$ in such loops. However, for the shortest loop ($L=50$\, Mm; c.f. Figure 1a and 1b), the analytical approximation deviates from the numerical solution in the regime of low background density and high background temperature ($d \approx 1$).
 \begin{figure}
   \centerline{\hspace*{0.03\textwidth}
               \includegraphics[width=0.56\textwidth,clip=]{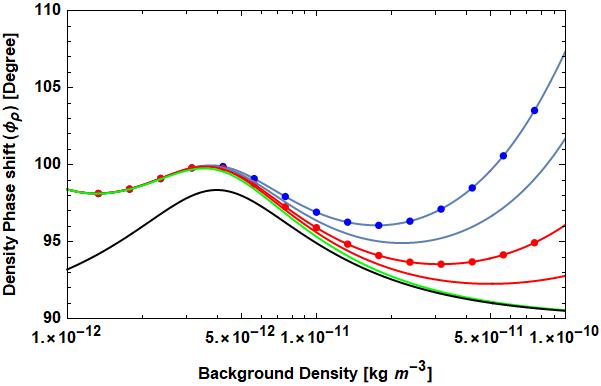}
               \hspace*{0.06\textwidth}
               \includegraphics[width=0.56\textwidth,clip=]{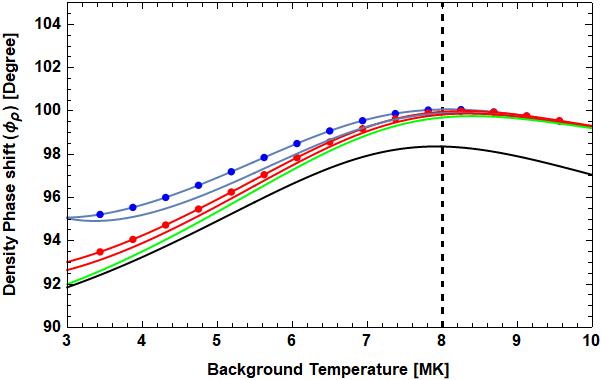}
              }
     \vspace{-0.35\textwidth}   
     \centerline{\Large \bf     
      \hspace{-0.17 \textwidth}  \color{black}{\footnotesize{(a)}}
      \hspace{0.58\textwidth}  \color{black}{\footnotesize{(b)}}
         \hfill}
     \vspace{0.31\textwidth}    
   \centerline{\hspace*{0.015\textwidth}
               \includegraphics[width=0.56\textwidth,clip=]{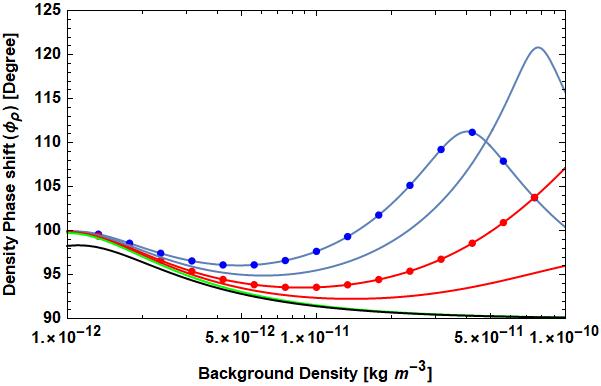}
               \hspace*{0.06\textwidth}
               \includegraphics[width=0.56\textwidth,clip=]{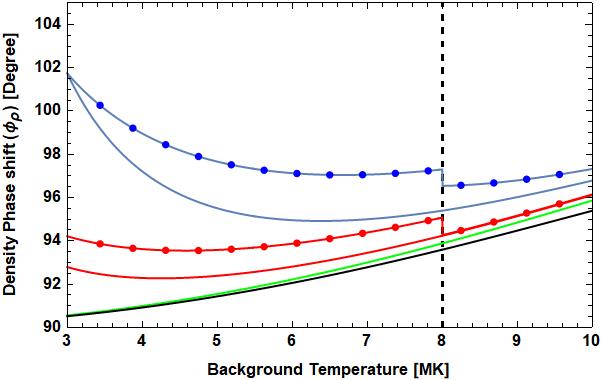}
              }
     \vspace{-0.35\textwidth}   
     \centerline{\Large \bf     
      \hspace{-0.17 \textwidth} \color{black}{\footnotesize{(c)}}
      \hspace{0.58\textwidth}  \color{black}{\footnotesize{(d)}}
         \hfill}
     \vspace{0.31\textwidth}    
   \centerline{\hspace*{0.015\textwidth}
               \includegraphics[width=0.56\textwidth,clip=]{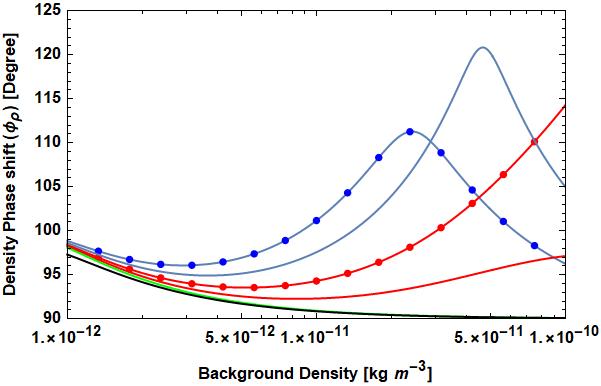}
               \hspace*{0.06\textwidth}
               \includegraphics[width=0.56\textwidth,clip=]{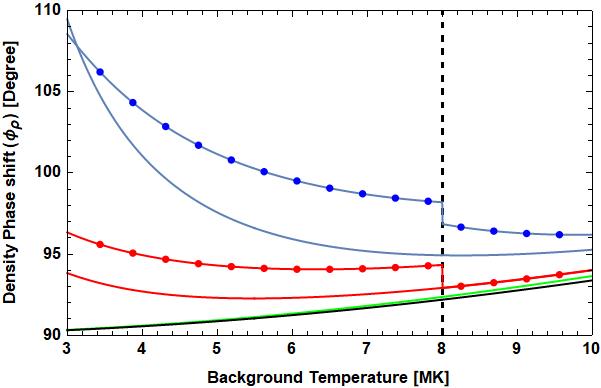}
              }
     \vspace{-0.35\textwidth}   
     \centerline{\Large \bf     
      \hspace{-0.17 \textwidth} \color{black}{\footnotesize{(e)}}
      \hspace{0.58\textwidth}  \color{black}{\footnotesize{(f)}}
         \hfill}
     \vspace{0.31\textwidth}    

   \centerline{\hspace*{0.015\textwidth}
               \includegraphics[width=0.56\textwidth,clip=]{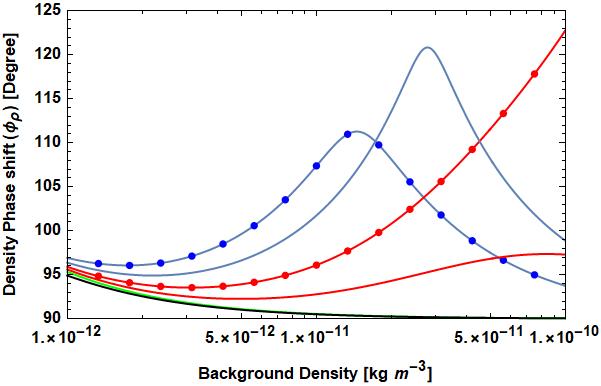}
               \hspace*{0.06\textwidth}
               \includegraphics[width=0.56\textwidth,clip=]{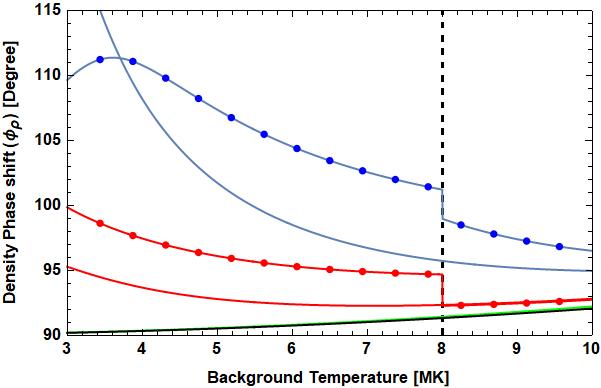}
              }
     \vspace{-0.35\textwidth}   
     \centerline{\Large \bf     
      \hspace{-0.17 \textwidth} \color{black}{\footnotesize{(g)}}
      \hspace{0.58\textwidth}  \color{black}{\footnotesize{(h)}}
         \hfill}
     \vspace{0.31\textwidth}    
     
\caption{Variation of $\phi_{\rho}$ for loop lengths of $L=50,180,300,500$\, Mm ({\it panels from top to bottom}). {\it Panels of left column} show the variation w.r.t. background density  at constant temperature of $T_0 = 5$\, MK whereas {\it panels of right column} show the variations w.r.t. background temperature at constant density of $\rho_{0} = 10^{-11}$\,kg\, ${\text m}^{-3}$. The combined effect of thermal conductivity, viscosity, and radiative losses (Model I) with heating function $a=-0.5, b=-3$ (or constant heating) is represented by the {\it blue (or red) solid curves}, the {\it green-solid curves} show the combined role of thermal conductivity and viscosity while the {\it black solid ones} are obtained by considering the effect of thermal conductivity only. The curves with the {\it blue (or red) circles} are for the case of radiative Model II with heating function $a=-0.5, b=-3$ (or constant heating).} 
 \label{F-4panels}
 \end{figure}
}
\subsection{Role of Compressive Viscosity}
{
In the present section we include the compressive viscous effects along with thermal conductivity in the basic MHD equations ($d, e \neq 0; r=0$). 

Figures 2 \,--\, 4 show the dependence of density phase shift [$\phi_{\rho}$], temperature phase [$\phi_{\rm T}$] shift, and phase difference [$\Delta \phi$] respectively on background density and temperature for loop lengths of $L = 50, 180, 300, 500$\, Mm. Note that the left panels in each figure are plotted with respect to the background density at constant temperature of $T_0 = 5$\, MK while the right panels in each figure are plotted with respect to background temperature considering a constant density of $10^{-11}$\, kg\, ${\text m}^{-3}$. In each panel the combined effect of thermal conductivity and compressive viscosity is represented by green-solid curves. For most of the panels in Figures 2 \,--\, 4 we see that the  green-solid curves coincide with the black-solid curves which are plotted considering the effect of thermal conductivity only. This suggests that the role of compressive viscosity on phase shifts is insignificant for almost the entire parametric space of loops considered in the study, which is expected from the small values of the viscous ratio (c.f. Table 1). Although we find that compressive viscosity increases the density phase shift slightly for low background densities and high background temperatures in the case of the shortest loop of $50$\, Mm (c.f. Panels 2a and 2b). 

In next section we discuss the phase shifts by including radiative losses with a constant heating into our model.
 \begin{figure}
   \centerline{\hspace*{0.03\textwidth}
               \includegraphics[width=0.56\textwidth,clip=]{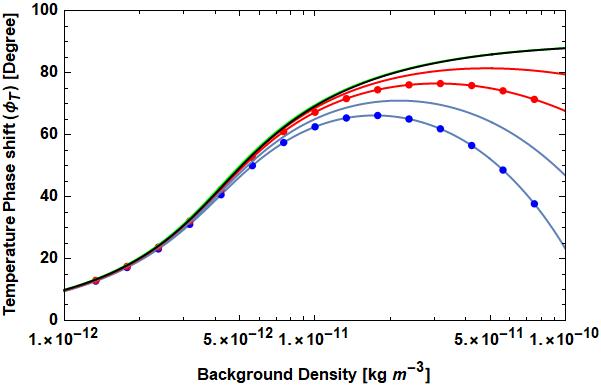}
               \hspace*{0.06\textwidth}
               \includegraphics[width=0.56\textwidth,clip=]{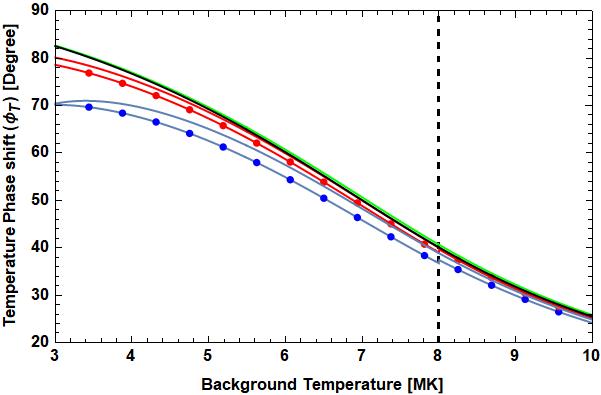}
              }
     \vspace{-0.35\textwidth}   
     \centerline{\Large \bf     
      \hspace{-0.17 \textwidth}  \color{black}{\footnotesize{(a)}}
      \hspace{0.58\textwidth}  \color{black}{\footnotesize{(b)}}
         \hfill}
     \vspace{0.310\textwidth}    
   \centerline{\hspace*{0.015\textwidth}
               \includegraphics[width=0.56\textwidth,clip=]{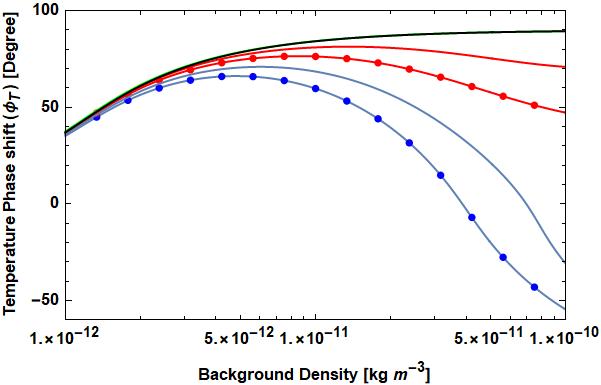}
               \hspace*{0.06\textwidth}
               \includegraphics[width=0.56\textwidth,clip=]{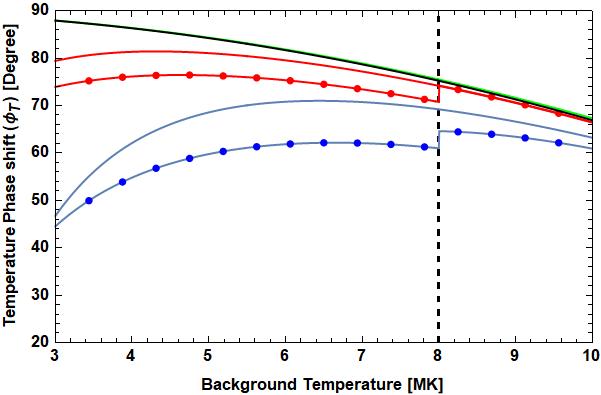}
              }
     \vspace{-0.35\textwidth}   
     \centerline{\Large \bf     
      \hspace{-0.17 \textwidth} \color{black}{\footnotesize{(c)}}
      \hspace{0.58\textwidth}  \color{black}{\footnotesize{(d)}}
         \hfill}
     \vspace{0.310\textwidth}    
   \centerline{\hspace*{0.015\textwidth}
               \includegraphics[width=0.56\textwidth,clip=]{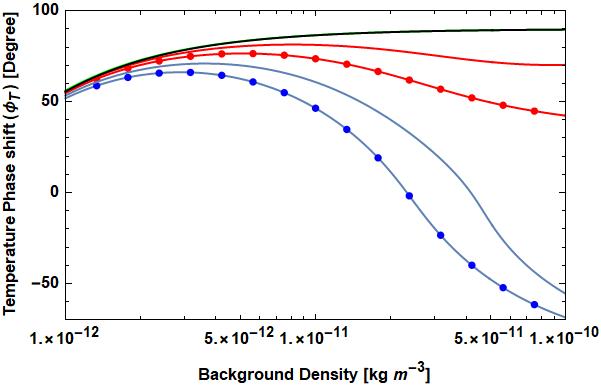}
               \hspace*{0.06\textwidth}
               \includegraphics[width=0.56\textwidth,clip=]{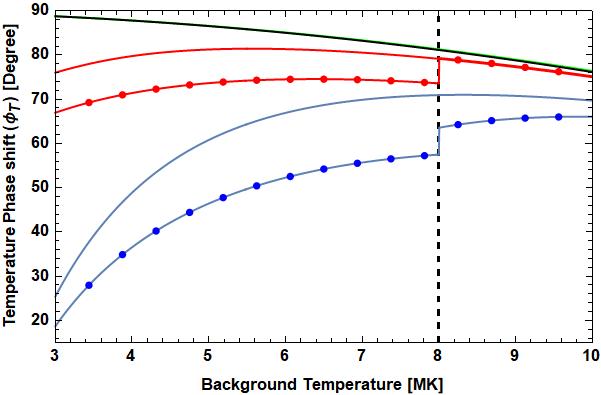}
              }
     \vspace{-0.35\textwidth}   
     \centerline{\Large \bf     
      \hspace{-0.17 \textwidth} \color{black}{\footnotesize{(e)}}
      \hspace{0.58\textwidth}  \color{black}{\footnotesize{(f)}}
         \hfill}
     \vspace{0.31\textwidth}    
   \centerline{\hspace*{0.015\textwidth}
               \includegraphics[width=0.56\textwidth,clip=]{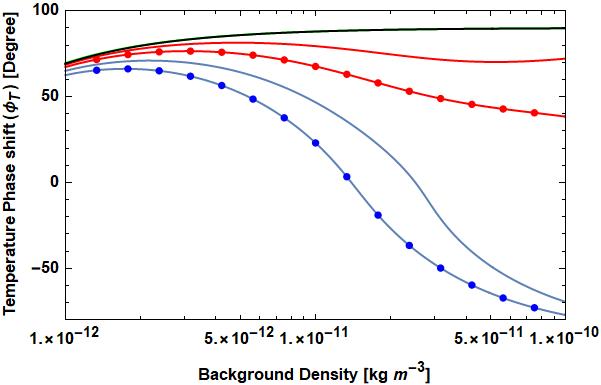}
               \hspace*{0.06\textwidth}
               \includegraphics[width=0.56\textwidth,clip=]{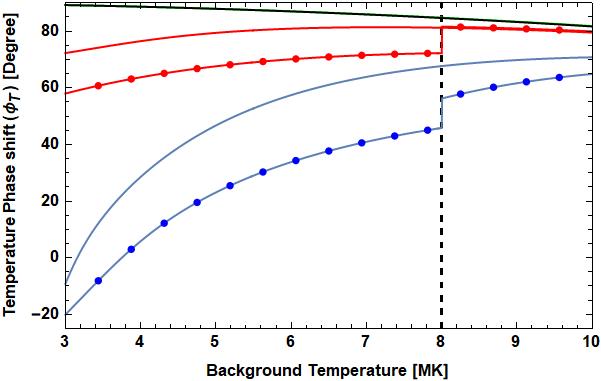}
              }
     \vspace{-0.35\textwidth}   
     \centerline{\Large \bf     
      \hspace{-0.17 \textwidth} \color{black}{\footnotesize{(g)}}
      \hspace{0.58\textwidth}  \color{black}{\footnotesize{(h)}}
         \hfill}
     \vspace{0.31\textwidth}    
     
\caption{Variation of $\phi_{\rm T}$ for loop lengths of $L=50,180,300,500$\, Mm ({\it panels from top to bottom}). {\it Panels of left column} show the variation w.r.t. background density  at constant temperature of $T_0 = 5$\, MK whereas {\it panels of right column} show the variations w.r.t. background temperature at constant density of $\rho_{0} = 10^{-11}$\,kg\, ${\text m}^{-3}$. The combined effect of thermal conductivity, viscosity, and radiative losses (Model I) with heating function $a=-0.5, b=-3$ (or constant heating) is represented by the {\it blue (or red) solid curves}, the {\it green-solid curves} show the combined role of thermal conductivity and viscosity while the {\it black solid ones} are obtained by considering the effect of thermal conductivity only. The curves with the {\it blue (or red) circles} are for the case of radiative Model II with heating function $a=-0.5, b=-3$ (or constant heating). Note that here {\it green-solid curves} completely overlap {\it black-solid ones} in most of the panels (see text for detail).}
 \label{F-4panels}
 \end{figure}
}
\subsection{Role of Radiative Losses with Constant Background Heating per Unit Mass} 
{We further include the radiative damping into our MHD model, apart from thermal conductivity and compressive viscosity. For the case of constant background heating we consider the free parameters of the heating function [$a$, $b$] to be simultaneously zero. 

The role of radiative effects with a constant heating is studied by solving the dispersion relation (Equation 16) and plotting the phase shifts for loop lengths from $L = 50$ \,--\, $500$\, Mm. The red-solid curves in each panel of Figures 2\,--\,4 show the solutions obtained by including the radiative losses (Model I) in the MHD model. Further, the red-solid curves with red circles are for the case of radiative function Model II given in Equation 10. In this case the index value abruptly changes from $\alpha = 1/3$ to $\alpha = -1$ at the breakpoint of $T_0 = 8$\, MK. We find that the use of this step-wise radiative-loss function leads to a significant deviation in phase shifts from that of constant $\alpha = -0.5$ especially for the higher loop lengths ($r\gg 1$). In the case of lower loop length ($L = 50$\, Mm) the approximation $\alpha =-0.5$ holds comparatively better for the entire range of background temperatures considered; however, at higher densities a significant deviation is still observed. Furthermore, both of the radiative-cooling models (with constant background heating) lead to exactly the same value of phase shifts in the temperature range of $T_0 = 8$ \,--\, $10$\, MK at all loop lengths. The overall radiative effect is observed on all the loops considered. Interestingly, for the largest loop length case ($L=500$\, Mm), at higher background densities the consideration of radiative damping decreases the density phase shift in comparison to the case when only thermal conductivity is considered (c.f. Figure 2g). The inclusion of radiative effects leads to an increment in the phase difference [$\Delta \phi$] and this increment is slightly more in the regime of high background density and low background temperature ($r \gg 1$).
 \begin{figure} 
   \centerline{\hspace*{0.03\textwidth}
               \includegraphics[width=0.56\textwidth,clip=]{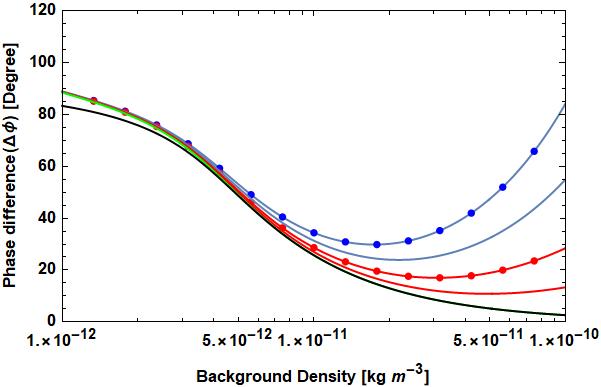}
               \hspace*{0.06\textwidth}
               \includegraphics[width=0.56\textwidth,clip=]{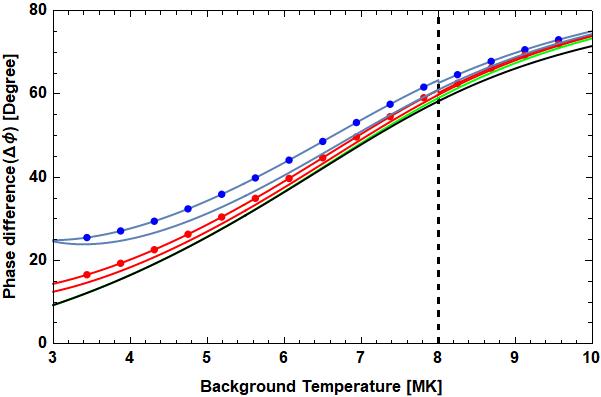}
              }
     \vspace{-0.35\textwidth}   
     \centerline{\Large \bf     
      \hspace{-0.17 \textwidth}  \color{black}{\footnotesize{(a)}}
      \hspace{0.58\textwidth}  \color{black}{\footnotesize{(b)}}
         \hfill}
     \vspace{0.31\textwidth}    
   \centerline{\hspace*{0.015\textwidth}
               \includegraphics[width=0.56\textwidth,clip=]{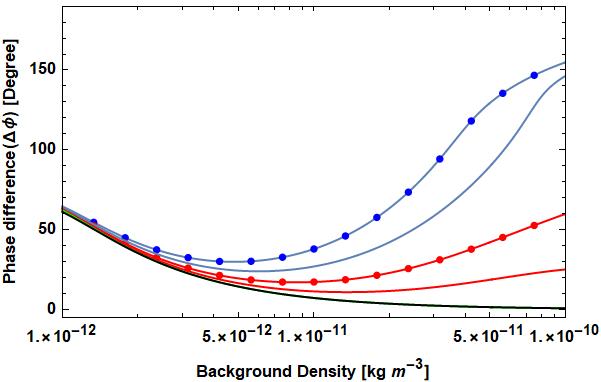}
               \hspace*{0.06\textwidth}
               \includegraphics[width=0.56\textwidth,clip=]{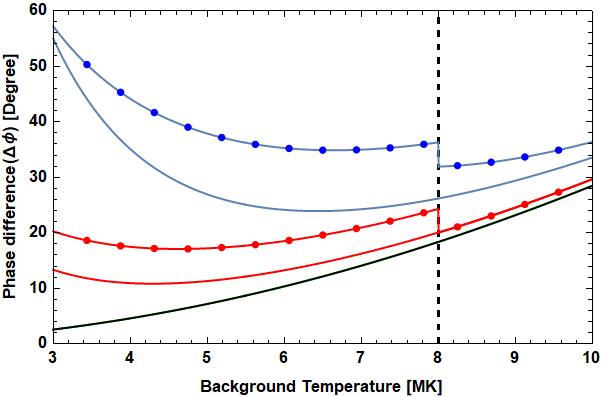}
              }
     \vspace{-0.35\textwidth}   
     \centerline{\Large \bf     
      \hspace{-0.17 \textwidth} \color{black}{\footnotesize{(c)}}
      \hspace{0.58\textwidth}  \color{black}{\footnotesize{(d)}}
         \hfill}
     \vspace{0.31\textwidth}    
   \centerline{\hspace*{0.015\textwidth}
               \includegraphics[width=0.56\textwidth,clip=]{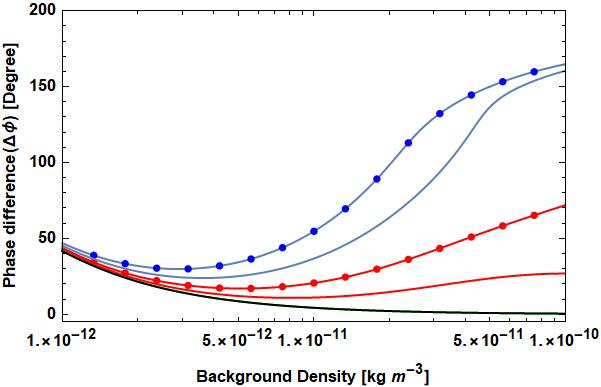}
               \hspace*{0.06\textwidth}
               \includegraphics[width=0.56\textwidth,clip=]{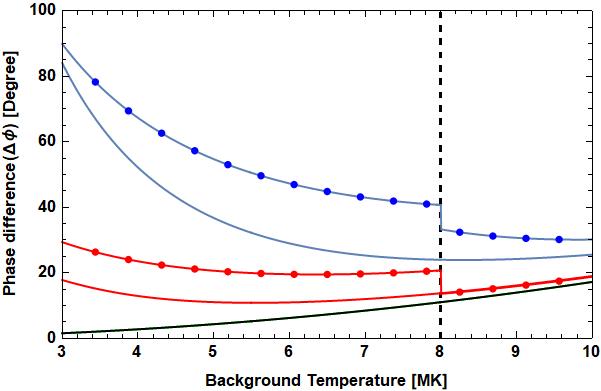}
              }
     \vspace{-0.35\textwidth}   
     \centerline{\Large \bf     
      \hspace{-0.17 \textwidth} \color{black}{\footnotesize{(e)}}
      \hspace{0.58\textwidth}  \color{black}{\footnotesize{(f)}}
         \hfill}
     \vspace{0.31\textwidth}    
   \centerline{\hspace*{0.015\textwidth}
               \includegraphics[width=0.56\textwidth,clip=]{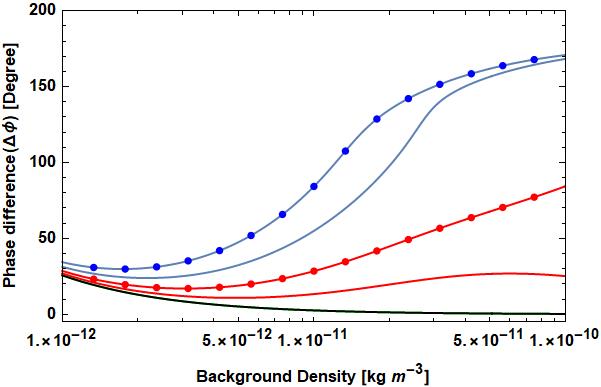}
               \hspace*{0.06\textwidth}
               \includegraphics[width=0.56\textwidth,clip=]{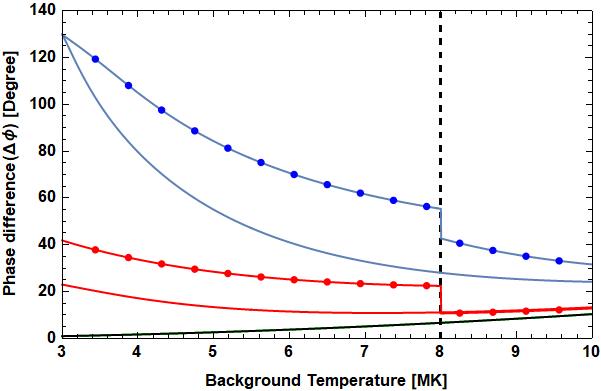}
              }
     \vspace{-0.35\textwidth}   
     \centerline{\Large \bf     
      \hspace{-0.17 \textwidth} \color{black}{\footnotesize{(g)}}
      \hspace{0.58\textwidth}  \color{black}{\footnotesize{(h)}}
         \hfill}
     \vspace{0.30\textwidth}    
     
\caption{Variation of $\Delta \phi$ for loop lengths of $L=50,180,300,500$\, Mm ({\it panels from top to bottom}). {\it Panels of left column} show the variation w.r.t. background density  at constant temperature of $T_0 = 5$\, MK whereas {\it panels of right column} show the variations w.r.t. background temperature at constant density of $\rho_{0} = 10^{-11}$\,kg\, ${\text m}^{-3}$. The combined effect of thermal conductivity, viscosity, and radiative losses (Model I) with heating function $a=-0.5, b=-3$ (or constant heating) is represented by the {\it blue (or red) solid curves}, the {\it green-solid curves} show the combined role of thermal conductivity and viscosity while the {\it black solid ones} are obtained by considering the effect of thermal conductivity only. The curves with the {\it blue (or red) circles} are for the case of radiative Model II with heating function $a=-0.5, b=-3$ (or constant heating). Note that here {\it green-solid curves} completely overlap {\it black-solid ones} in most of the panels (see text for detail).}  \label{F-4panels}
 \end{figure}
  
}
\subsection{Role of Density and Temperature Dependent Coronal Heating Function} 
{
In this section we study the general dispersion relation including all the damping effects of the MHD model described in Section 2, i.e. Equations 1 \,--\, 4. We will focus on the heating function given with
\begin{equation}
    a = -0.5 \,\,\,,\,\,\, b = -3.
\end{equation}
The heating--cooling misbalance (with different choices of $a$ and $b$) may lead to damped, undamped, or growing oscillations \citep{2019A&A...628A.133K}. The choice of $a=-0.5$ and $b=-3$ leads to an enhanced damping relative to other dissipation mechanisms and also provides an oscillation quality factor (ratio of decay time and period) in the range of $1$ \,--\, $2$ consistent with SUMER observations of standing slow-mode waves.

The solid blue curves in Figures 2 \,--\, 4 correspond to the case considering all the dissipative effects described in the model in Section 2. The solid curves with blue-circles are plotted using the radiative Model II (Equation 10). The use of step-wise radiative function leads to a significant deviation in phase shifts compared to the case of $\alpha = -0.5$ (Model I) for all loop lengths considered. This is because the consideration of different radiation functions alters the dispersion relation and eventually the frequency $\omega$, which leads to a change in the thermal-misbalance influence on the phase shifts. We find that larger loop lengths have a more significant effect from the considered heating function as compared to shorter loops. For the shortest loop case ($L=50$\, Mm), the heating function leads to a significant increase in the density phase shift and lowers the temperature phase shift in the regime of higher background density (c.f. top panels of Figures 2 and 3). For $L=180$\, Mm, it has a significant effect in the loops of high background density or low background temperature. Figure 4c shows that the inclusion of the given heating function sharply increases the phase difference to almost $\Delta \phi \approx 150^\circ$ for bulky loop with $\rho_0 = 10^{-10}$\, kg\, ${\text m}^{-3}$. For large loop lengths ($L=300,500$\, Mm), we observe that the density phase shift reaches a maximum value for some intermediate loop density when the heating function is considered (c.f. Figures 2e and 2g). Moreover, the role of heating function is very drastic for the case of higher background density or lower background temperature and we observe that the phase difference is significantly larger than that for the case with constant heating in this regime (c.f. Figure 4).

The role of different dissipative effects on the polytropic index is discussed in the following section.
}

\subsection{Analysis of the Polytropic Index} 
{
We study the polytropic index using the linear MHD model and derive a theoretical expression for it under the effect of all the dissipative mechanisms. Note that the further analysis is similar to that done by \citet{2021SoPh..296..105P} for propagating waves, and we provide a follow-up of their work regarding the nature of the polytropic index for the case of standing slow-mode waves. The linearized energy equation is given as,
\begin{equation}
    \frac{\partial T_1}{\partial t} = (\gamma-1)\frac{\partial \rho_1}{\partial t} + \gamma d  \left(  \frac{\partial^2 T_1}{\partial z^2} \right) + \gamma r(b-\alpha)T_1 + \gamma r(a-1)\rho_1.
\end{equation}
Further since we have
\begin{align}
    \rho_1 &= \hat{\rho}_1({\text e}^{{\rm i}(kz-\omega t)} + {\text e}^{{\rm i}(-kz-\omega t)}),\\
    T_1 &= \hat{T}_1 ({\text e}^{{\rm i}(kz-\omega t - \Delta \phi)}+{\text e}^{{\rm i}(-kz-\omega t - \Delta \phi)} ).
\end{align}
Substituting the above expressions in the energy equation we get
\begin{equation}
    {\text e}^{-{\rm i}\Delta \phi}(\gamma dk^2 - \gamma r(b - \alpha) - {\rm i}\omega )\hat{T}_1 = (\gamma r(a-1) - {\rm i}\omega (\gamma-1))\hat{\rho}_1.
\end{equation}
Separating the imaginary and real components we have
\begin{align}
    [-(\gamma d k^2 + \omega_{\rm i} - \gamma r(b-\alpha))\sin \Delta \phi - \omega_{\rm r} \cos \Delta \phi]\,\hat{T}_1 &= -\omega_{\rm r}(\gamma-1)\hat{\rho}_1,\\
    [(\gamma d k^2 + \omega_{\rm i} - \gamma r(b-\alpha))\cos \Delta \phi - \omega_{\rm r} \sin \Delta \phi]\,\hat{T}_1 &= ((\gamma-1)\omega_{\rm i} + \gamma r (a-1))\hat{\rho}_1.
\end{align}
We multiply Equation 50 with $\cos \Delta \phi$ and Equation 51 with $\sin \Delta \phi$ then finally add the two equations to obtain
\begin{multline}
    \hat{T}_1 = \left[(\gamma-1)\cos \Delta \phi - \frac{\omega_{\rm i}}{\omega_{\rm r}}(\gamma-1)\sin \Delta \phi - \frac{\gamma r}{\omega_{\rm r}}(a-1)\sin \Delta \phi \right]\hat{\rho}_1 \equiv \\ (\gamma_{\rm eff}-1)\hat{\rho}_1,
\end{multline}
where $\gamma_{\rm eff}$ is defined under the polytropic assumption (i.e. $p \sim \rho^{\gamma_{\rm eff}}$).\newline
Thus we have 
\begin{equation}
    \gamma_{\rm eff}-1 = (\gamma-1)\cos \Delta \phi - \frac{\omega_{\rm i}}{\omega_{\rm r}}(\gamma-1)\sin \Delta \phi - \frac{\gamma r}{\omega_{\rm r}}(a-1)\sin \Delta \phi. 
\end{equation}
If only thermal conductivity is present then 
\begin{equation}
    \gamma_{\rm eff}-1 = \left(1 - \frac{\omega_{\rm i}}{\omega_{\rm r}}\tan \Delta \phi\right)(\gamma-1)\cos \Delta \phi.
\end{equation}
Substituting the expression in Equation 43 into Equation 54 we get
\begin{equation}
    \gamma_{\rm eff}-1 = \frac{(\gamma-1)\cos \Delta \phi}{1 + \gamma d \omega_{\rm i}k^2/(\omega_{\rm r}^2 + \omega_{\rm i}^2)},
\end{equation}
Under the weak damping assumption, $\omega_{\rm i} \approx 0$, we obtain 
\begin{equation}
\gamma_{\rm eff}-1=(\gamma-1)\cos(\Delta\phi).
\end{equation}
which is consistent with that given by \citet{2015ApJ...811L..13W,2018ApJ...860..107W}.

 \begin{figure}   
   \centerline{\hspace*{0.03\textwidth}
               \includegraphics[width=0.56\textwidth,clip=]{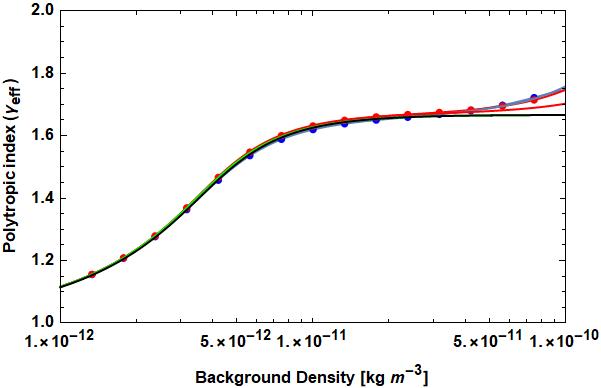}
               \hspace*{0.06\textwidth}
               \includegraphics[width=0.56\textwidth,clip=]{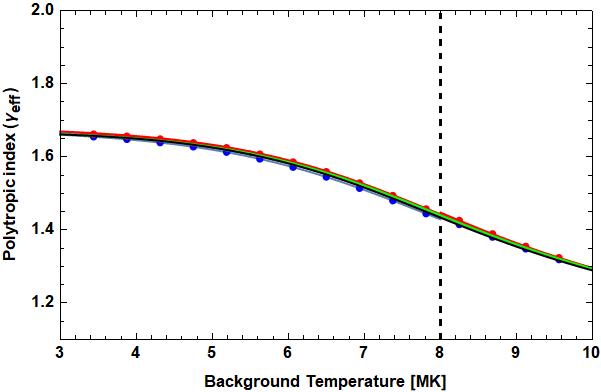}
              }
     \vspace{-0.35\textwidth}   
     \centerline{\Large \bf     
      \hspace{-0.17 \textwidth}  \color{black}{\footnotesize{(a)}}
      \hspace{0.58\textwidth}  \color{black}{\footnotesize{(b)}}
         \hfill}
     \vspace{0.31\textwidth}    
   \centerline{\hspace*{0.015\textwidth}
               \includegraphics[width=0.56\textwidth,clip=]{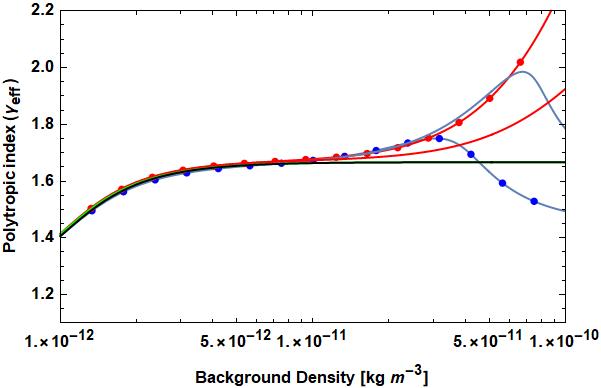}
               \hspace*{0.06\textwidth}
               \includegraphics[width=0.56\textwidth,clip=]{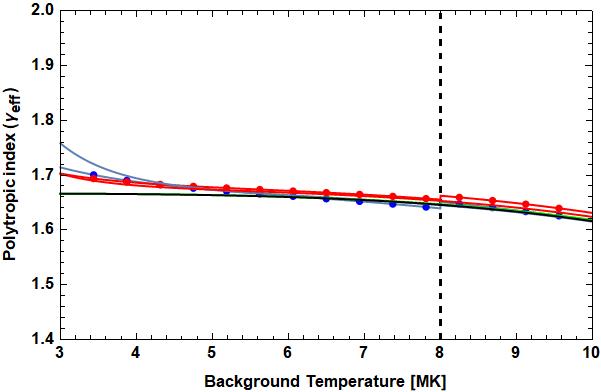}
              }
     \vspace{-0.35\textwidth}   
     \centerline{\Large \bf     
      \hspace{-0.17 \textwidth} \color{black}{\footnotesize{(c)}}
      \hspace{0.58\textwidth}  \color{black}{\footnotesize{(d)}}
         \hfill}
     \vspace{0.31\textwidth}    
   \centerline{\hspace*{0.015\textwidth}
               \includegraphics[width=0.56\textwidth,clip=]{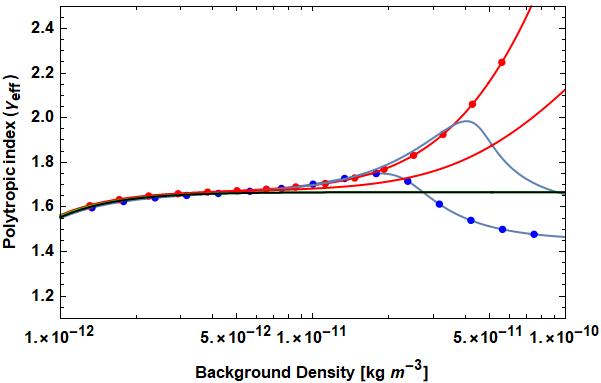}
               \hspace*{0.06\textwidth}
               \includegraphics[width=0.56\textwidth,clip=]{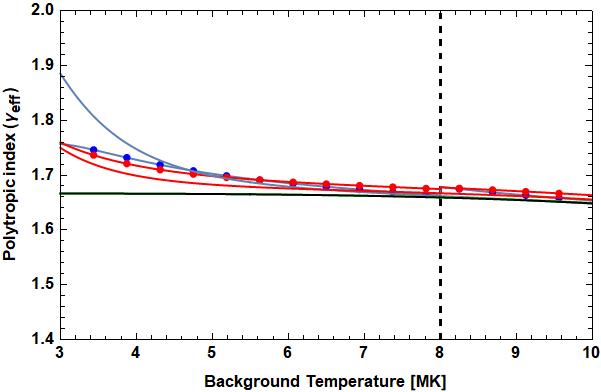}
              }
     \vspace{-0.35\textwidth}   
     \centerline{\Large \bf     
      \hspace{-0.17 \textwidth} \color{black}{\footnotesize{(e)}}
      \hspace{0.58\textwidth}  \color{black}{\footnotesize{(f)}}
         \hfill}
     \vspace{0.31\textwidth}    
   \centerline{\hspace*{0.015\textwidth}
               \includegraphics[width=0.56\textwidth,clip=]{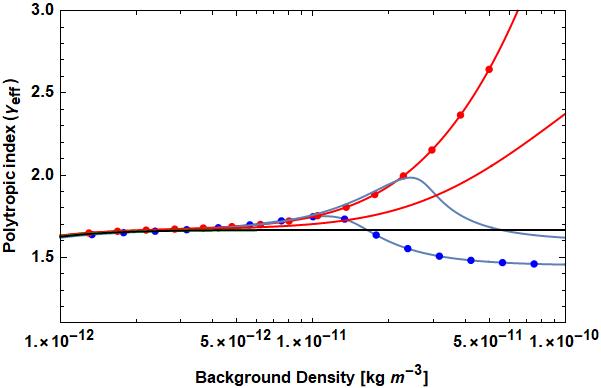}
               \hspace*{0.06\textwidth}
               \includegraphics[width=0.56\textwidth,clip=]{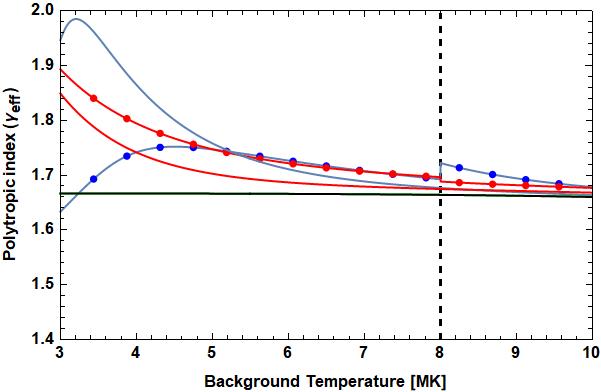}
              }
     \vspace{-0.35\textwidth}   
     \centerline{\Large \bf     
      \hspace{-0.17 \textwidth} \color{black}{\footnotesize{(g)}}
      \hspace{0.58\textwidth}  \color{black}{\footnotesize{(h)}}
         \hfill}
     \vspace{0.31\textwidth}    
     
\caption{Variation of $\gamma_{\rm eff}$ for loop lengths of $L=50,180,300,500$\, Mm ({\it panels from top to bottom}). {\it Panels of left column} show the variation w.r.t. background density  at constant temperature of $T_0 = 5$\, MK whereas {\it panels of right column} show the variations w.r.t. background temperature at constant density of $\rho_{0} = 10^{-11}$\,kg\, ${\text m}^{-3}$. The combined effect of thermal conductivity, viscosity, and radiative losses (Model I) with heating function $a=-0.5, b=-3$ (or constant heating) is represented by the {\it blue (or red) solid curves}, the {\it green-solid curves} show the combined role of thermal conductivity and viscosity while the {\it black solid ones} are obtained by considering the effect of thermal conductivity only. The curves with the {\it blue (or red) circles} are for the case of radiative Model II with heating function $a=-0.5, b=-3$ (or constant heating). Note that here {\it green-solid curves} completely overlap {\it black-solid ones} in most of the panels (see text for detail).}
 \label{F-4panels}
 \end{figure}

Figure 5 shows the curves of polytropic index with respect to background density and temperature. The top panels correspond to the shortest loop of $L=50$\, Mm, while the bottom panels are for $L=500$\, Mm. It is clearly visible from Figure 5 that the compressive viscosity has negligible effect on the polytropic index since the green-solid curves completely overlap the black-solid ones. Keep in mind that here also the variations with background density are shown by keeping a constant temperature of $T_0 = 5$ \,MK and the variations with background temperature are shown for $\rho_0 = 10^{-11}$\,kg\, ${\text m}^{-3}$. In the case of the shortest loop of $50$\, Mm, for most of the background densities, the polytropic index is close to 1.66 (c.f. Figures 5a and 5b) but it reduces to $\approx 1.1$ for low background density ($\rho_0 \approx 10^{-12}$\, kg\, ${\text m}^{-3}$), and it gradually reduces to $\approx 1.3$ at higher background temperatures ($T_0 = 10$\, MK), mainly due to thermal conductivity. That is, the radiative damping and heating\,--\,cooling misbalance have only a weak effect on $\gamma_{\rm eff}$ in this shortest loop ($L=50$\, Mm). Moreover, the radiative-cooling Model II (curves with filled red and blue circles corresponding to the cases of constant heating and specific heating function) also has negligible influence at this loop length and the approximation with constant $\alpha = -0.5$ holds well. We find that for the cases with $L=180,300,500$\, Mm the polytropic index is close to the classical value ($\gamma_{\rm eff}=5/3$) for most of the equilibrium densities ($\rho_0 > 10^{-11}$\, kg\,${\text m}^{-3}$) and nearly all the temperatures ($T_0<10$\, MK) when considering thermal conduction alone. This suggests that these loops correspond to the weak thermal-conduction regime (i.e. $d\ll0.1$, see Table 1). Note that the damping rate of thermal conduction reaches the maximum value at $d \approx 0.1$ for the fundamental mode. \citep[c.f.][]{2003A&A...408..755D,2021SSRv..217...34W}. In the case of $L=180$\, Mm, we observe that the presence of a constant background heating ($a=0$, $b=0$) monotonically increases the polytropic index with increasing background density (red-solid curves with and without red circles)  and in contrast for the case of $a=-0.5$ and $b=-3$ the $\gamma_{\rm eff}$ reaches a peak value of $\approx 2.0$ (blue-solid curve, Model I) and $\approx 1.7$ (Model II). Furthermore, the $\gamma_{\rm eff}$ slightly increases from its classical value at lower background temperatures when radiative cooling (Model I and II) with constant background heating is considered (c.f. Figure 5d). For the radiative Model II we find that there is a significant deviation in the value of polytropic index at higher background densities when constant heating or the heating function is considered (c.f. Figure 5c). In the case of large loop lengths we observe similar results as for intermediate loop length ($L=180$\, Mm), however, we find that the rate of increase of $\gamma_{\rm eff}$ with background density is larger when constant heating is considered and the polytropic index monotonically increases to a comparatively higher values (c.f. Figures 5e and 5g). Moreover we find that in the case of considered heating function ($a=-0.5, b=-3$), the peak value of $\gamma_{\rm eff}$ is reached at comparatively lower background densities for large loop lengths. The $\gamma_{\rm eff}$ gradually reduces to $\approx 1.4$ for radiative Model II with the considered heating function at higher densities. In summary, the radiative effects are more significant for larger loop lengths (because of the radiative ratio $r \propto L$) and there is a sharp contrast in the nature of $\gamma_{\rm eff}$ between the cases of constant heating and specific heating function ($a=-0.5, b=-3$). We also find that for the loop length $L=500$\, Mm at lower background temperatures the step-wise radiative function reduces the $\gamma_{\rm eff}$ to $1.66$ whereas the approximation $\alpha = -0.5$ increases the polytropic index to nearly 2 (c.f. Figure 5h).

}
\subsection{Comparison of Phase Shifts and Polytropic Index in Standing and Propagating Slow-Mode Waves}
{
In the present section we emphasize on the key differences in the linear modelling of the phase shifts in propagating and standing slow-mode waves. The propagating slow-mode waves in \citet{2021SoPh..296..105P} were considered with a Fourier wave solution of the form
\begin{equation}
    v = \hat{v}_1{\text e}^{{\rm i}(kz-\omega t)},
\end{equation}
where $k = k_{\rm r} + {\rm i} k_{\rm i} = k_{\rm m}{\text e}^{{\rm i}\phi}$ and $\omega$ is real. In their work the phase shift in space of the density [$z_\rho$] and temperature [$z_T$] perturbations with respect to velocity perturbations was calculated for the warm coronal loops in the equilibrium temperature range of $T_0 = 1$ \,--\, $2$\, MK. For the considered range of temperatures and densities it was found that the role of a specific heating function (with $a=-0.5$ and $b=-3$) was more significant for the loops with higher densities and lower temperatures, a condition that is favorable for wave dissipation by radiative losses (see Equations 13 and 14). Also the phase difference [$\Delta \phi$] in space between the density and temperature perturbations was found to increase drastically with inclusion of heating function at very low temperatures. \citet{2021SoPh..296..105P} found that for the considered range of loops, the polytropic index [$\gamma_{\rm eff}$] reduces to $\approx 1.2$ for lower densities and this reduction is primarily due to the effect of thermal conductivity ($d \propto \frac{1}{\rho_0}$). The polytropic index was also found to slightly increase from its classical value at lower background temperatures when constant heating was considered ($a=0$, $b=0$) whereas it slightly reduced from its classical value when heating function with $a=-0.5$ and $b=-3$ was considered. 

For the standing slow-mode waves we considered a Fourier solution being the sum of oppositely propagating wave solutions
\begin{equation}
    v = \hat{v}_1 ({\text e}^{{\rm i}(kz-\omega t)} + {\text e}^{{\rm i}(-kz-\omega t)}),
\end{equation}
where $k$ = 2n$\pi$ is real and $\omega = \omega_{\rm r} + {\rm i} \omega_{\rm i}$. We calculated the phase shift in time of the temperature [$\phi_{\rm T}$] and density [$\phi_{\rho}$] perturbations and also found that in contrast to the case of propagating slow-mode waves there is also an additional $\frac{\pi}{2}$ phase shift in both space and time of the density and temperature perturbations with respect to velocity perturbations (c.f. Equations 24 and 28). In contrast to \citet{2021SoPh..296..105P}, the present work focuses on a different regime of loops with equilibrium temperatures from $T_0 = 3$ \,--\, $10$\, MK and density in the range of $10^{-12}$ \,--\, $10^{-10}$\, kg \,${\text m}^{-3}$. It is also important to note that for the standing slow-mode waves the overall effect of heating function (with $a=-0.5$ and $b=-3$) on $\phi_\rho$ and $\phi_{\rm T}$ is found to be strongly dependent on the loop length [$L$] as well. This is expected as the radiative ratio also depends on the loop length; in the case of propagating waves the effect should be dependent on the wave length (or wave period). We also calculated the phase difference in time [$\Delta \phi$] between the density and temperature perturbations and find that the effect of heating function ($a=-0.5, b=-3$) on phase shift dependent on temperature is similar in the two cases (standing and propagating waves), while its effect on phase shift dependent on density is distinctly different \citep{2021SoPh..296..105P}. In Section 3.5 we studied the nature of the polytropic index for the case of standing slow-mode waves and quite interestingly we find that with changing background density the $\gamma_{\rm eff}$ reaches a peak value for some intermediate density when heating function is considered (c.f. Figure 5). This sort of behaviour was not observed by \citet{2021SoPh..296..105P} for the case of propagating slow-mode waves since it occurs due to the additional term proportional to $\omega_{\rm i}$ in the expression for the polytropic index (Equation 53) for the standing waves. However we do find that at lower background densities $\gamma_{\rm eff}$ reduces from its classical value due to the dissipation by thermal conduction and a similar conclusion was inferred by \citet{2021SoPh..296..105P}. In the variation of polytropic index with background temperature and density, we find a decrease in $\gamma_{\rm eff}$ at lower background temperatures and higher background densities with the inclusion of the heating function ($a=-0.5, b=-3$) and radiative Model II. \citet{2021SoPh..296..105P} similarly found that for the propagating waves, the inclusion of heating function (with $a=-0.5$ and $b=-3$) does not always lead to an increased $\gamma_{\rm eff}$ and at lower temperatures there was a slight reduction in its value.
}

\section{Discussion and Conclusions}
{
The phase shifts of standing, slow MHD oscillations are an important aspect of coronal seismology, and in the present work we have used the linear MHD model to analyse this property of slow-mode waves in a detailed manner. We have systematically solved the dispersion relation (Equation 16) and provided theoretical expressions for the phase shifts of density [$\phi_\rho$] and temperature [$\phi_{\rm T}$] perturbations in standing, slow-mode waves. We performed a parametric study by considering loops with densities ranging from $\rho_0 = 10^{-12}$ \,--\, $10^{-10}$\, kg\, ${\text m}^{-3}$, temperatures from $T_0 = 3$ \,--\, $10$\, MK, and loop lengths in the range of $50$ \,--\, $500$\, Mm. We discussed the individual effect of all the damping mechanisms on phase shifts under sections 3.1, 3.2, 3.3, and 3.4. We found that the phase difference [$\Delta \phi$] is nearly independent of the compressive viscous damping for a majority of loops considered in the study. This result agrees with that obtained by \citet{2019ApJ...886....2W} using nonlinear 1D MHD modelling. The effect of radiative losses (Model I and II) with constant background heating is more significant in higher equilibrium density ($\rho_0 \approx 10^{-10}$\, kg\, ${\text m}^{-3}$) and lower equilibrium temperature loops ($T_0 \approx 3$\, MK). Furthermore, the inclusion of heating--cooling misbalance with a specific heating function ($a=-0.5, b=-3$) drastically  increases the phase difference in comparison to the case of constant heating for loops with higher background densities or lower background temperatures (a condition favorable to the role of radiative losses in wave dissipation, i.e. the case with larger radiative ratio $r$).

We also derived a general expression of polytropic index [$\gamma_{\rm eff}$] using our comprehensive model (c.f. Equation 53) and we found that in linear MHD model the compressive viscosity has negligible effect on the polytropic index. The effect of radiative losses (Model I and II) with constant heating on $\gamma_{\rm eff}$ is very different from the case when a specific heating function ($a=-0.5$ and $b=-3$) is considered. In the case of constant heating, $\gamma_{\rm eff}$ monotonically increases from a value of $1.66$ as we move to higher background densities, while in contrast for the case of specific heating function it takes a peak value with increasing density and reduces to values lower than $1.66$ at higher densities. The radiative Model II leads to a contrasting change in polytropic-index values at lower background temperatures and higher background densities compared to Model I for the loop length of $L=500$\, Mm.
 \begin{figure}
   \centerline{\hspace*{0.03\textwidth}
               \includegraphics[width=0.63\textwidth,clip=]{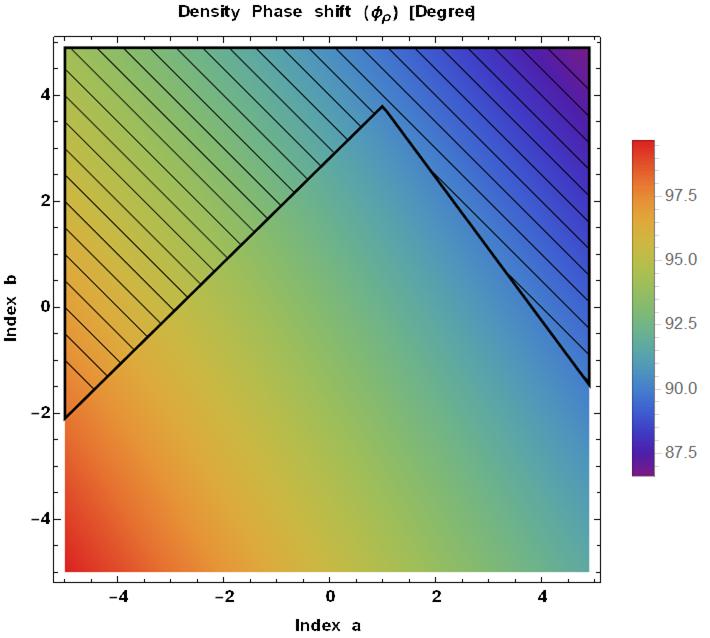}
               \hspace*{0.06\textwidth}
               \includegraphics[width=0.63\textwidth,clip=]{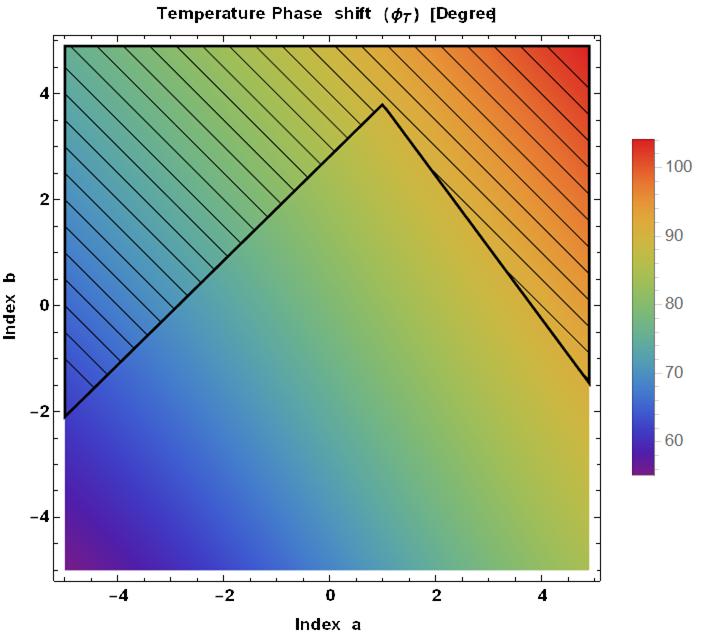}
              }
     \vspace{-0.35\textwidth}   
     \centerline{\Large \bf     
      \hspace{-0.25 \textwidth}  \color{black}{\footnotesize{(a)}}
      \hspace{0.67\textwidth}  \color{black}{\footnotesize{(b)}}
         \hfill}
     \vspace{0.315\textwidth}    
   \centerline{\hspace*{0.015\textwidth}
               \includegraphics[width=0.63\textwidth,clip=]{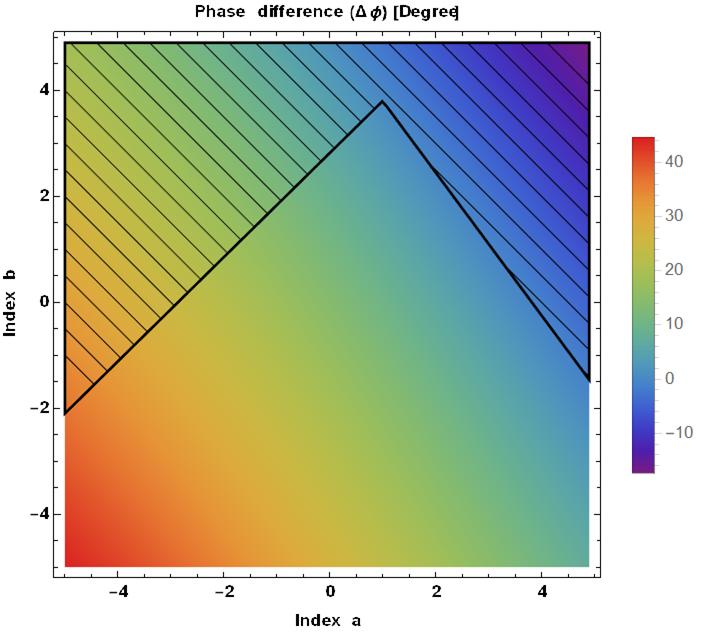}
               \hspace*{0.06\textwidth}
               \includegraphics[width=0.63\textwidth,clip=]{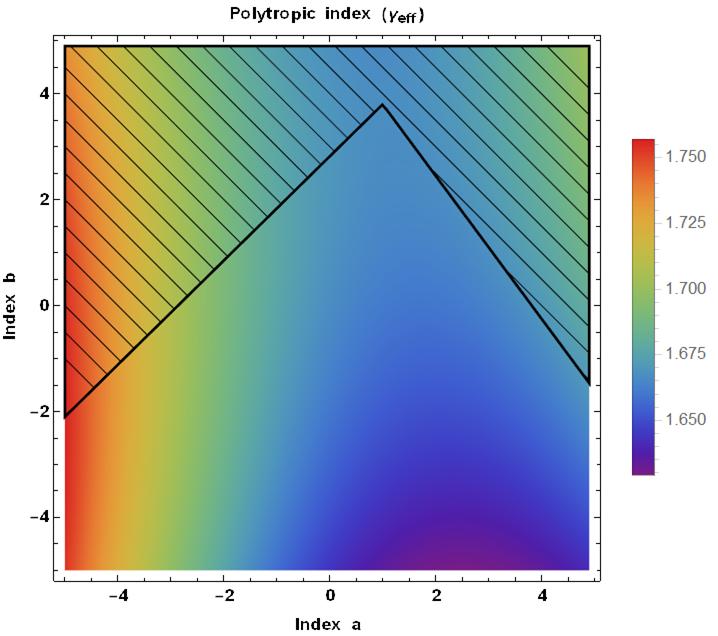}
              }
     \vspace{-0.35\textwidth}   
     \centerline{\Large \bf     
      \hspace{-0.25 \textwidth} \color{black}{\footnotesize{(c)}}
      \hspace{0.67\textwidth}  \color{black}{\footnotesize{(d)}}
         \hfill}
     \vspace{0.315\textwidth}    

\caption{The variation of density phase shift ({\it Panel a}), temperature phase shift ({\it Panel b}), Phase difference ({\it Panel c}) and polytropic index ({\it Panel d}) of fundamental mode with respect to the free parameters of heating function ($a$,$b$) for radiative cooling Model I. The panels are plotted for a loop of $T_0 = 6.3$\, MK, $\rho_0 = 10^{-11}$\, kg\, ${\text m}^{-3}$, and loop length $L=180$\, Mm. The hatched regions in each panel correspond to the unstable solutions where $\omega_{\rm i} > 0$}
 \label{F-4panels}
 \end{figure}
 \begin{figure}  
   \centerline{\hspace*{0.03\textwidth}
               \includegraphics[width=0.63\textwidth,clip=]{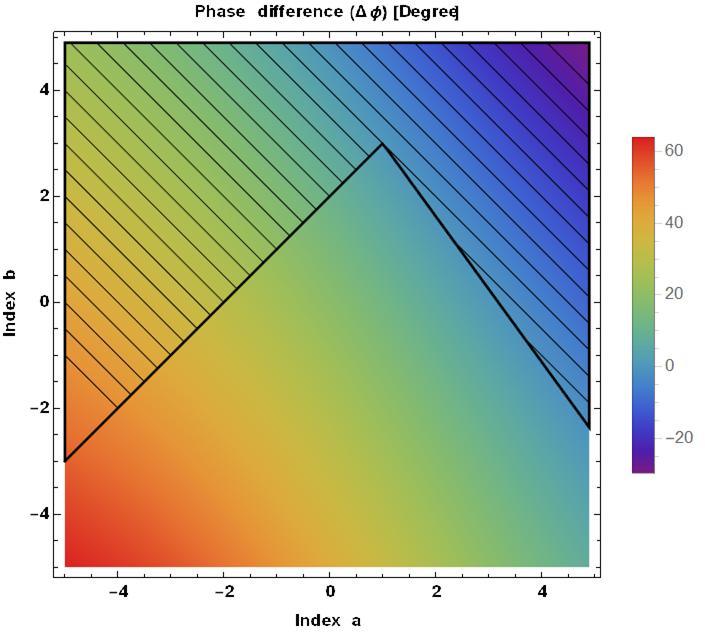}
               \hspace*{0.06\textwidth}
               \includegraphics[width=0.63\textwidth,clip=]{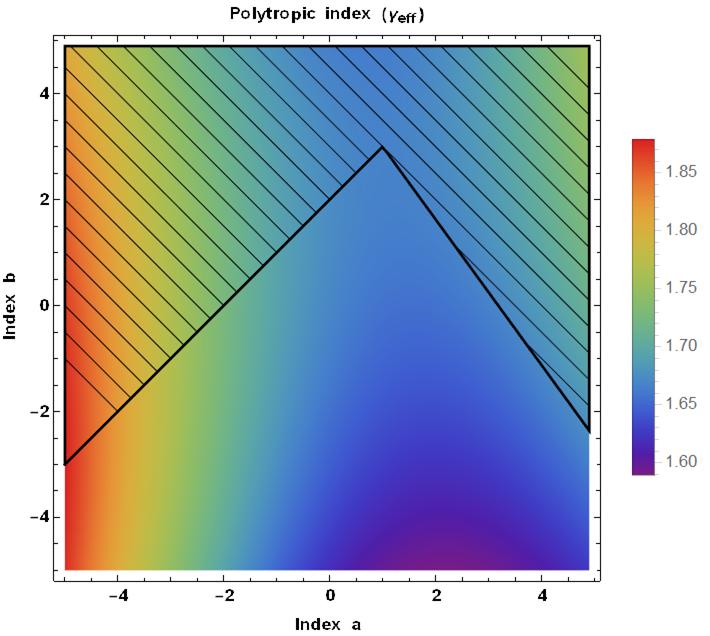}
              }
     \vspace{-0.35\textwidth}   
     \centerline{\Large \bf     
      \hspace{-0.25 \textwidth}  \color{black}{\footnotesize{(a)}}
      \hspace{0.67\textwidth}  \color{black}{\footnotesize{(b)}}
         \hfill}
     \vspace{0.315\textwidth}    

\caption{The variation of phase difference ({\it Panel a}) and polytropic index ({\it Panel b}) of fundamental mode with respect to the free parameters of heating function ($a$,$b$) for radiative cooling Model II. The panels are plotted for a loop of $T_0 = 6.3$\, MK, $\rho_0 = 10^{-11}$\, kg\, ${\text m}^{-3}$, and loop length $L=180$\, Mm. The hatched regions in each panel correspond to the unstable solutions where $\omega_{\rm i} > 0$}
 \label{F-4panels}
 \end{figure}

Since we have constrained our analysis until now to a guessed heating function ($a=-0.5, b=-3$), so in order to study the effect of different values of $a$- and $b$-parameters we look at the phase shifts for a specific loop with the parameters $T_0 = 6.3$\, MK, $\rho_0 = 10^{-11}$\, kg\, ${\text m}^{-3}$, and loop length $L=180$\, Mm, which are typical for SUMER loop oscillations \citep[e.g.][]{2007ApJ...656..598W}. For this loop the thermal ratio is $d=0.019$, viscous ratio $e=0.001$, and the radiative ratio $r=0.173$ (Model I). In Figure 6 we show the density plots depicting phase shifts and polytropic index for a range of free parameters $a$ and $b$ of the heating function using radiative Model I. The density phase shift  decreases for large values of $a$ and $b$ while the temperature phase shift increases for large $a$ and $b$. The shaded region in each panel corresponds to the unstable slow mode or thermal mode solutions ($\omega_{\rm i} > 0$) and the analysis is considered irrelevant for this shaded region. The $\gamma_{\rm eff}$ remains close to a value of $5/3$ for almost all the considered heating functions. The phase difference and polytropic index for the case of constant heating ($a=0$, $b=0$) are $\Delta \phi \approx 14^\circ$ and $\gamma_{\rm eff}=1.66$. For the case of heating--cooling misbalance with $a=-0.5$ and $b=-3$, the phase difference and polytropic index are $\Delta \phi \approx 24^\circ$ and $\gamma_{\rm eff} \approx 1.66$ (see Figures 4d and 5d). This result provides a validation for the assumed $\gamma_{\rm eff}=5/3$ in determination of magnetic-field strength by coronal seismology of standing slow-mode waves in hot coronal loops \citep{2007ApJ...656..598W}.

We have also used the radiative Model II to study the phase difference [$\Delta \phi$] and polytropic index for the loop studied in Figure 6. Figure 7 has shown the corresponding panels for phase difference and polytropic index. We find that for the given loop parameters there is a slight influence of the different radiative model, however the overall range of polytropic index [$\gamma_{\rm eff}$] remains close to the classical value of 1.66 as earlier. Here also the shaded region corresponds to the unstable slow or thermal mode solutions where the analysis is irrelevant.

\citet{2021SoPh..296..105P} also studied the polytropic index for different heating functions ($a$,$b$) but in a coronal loop with $T_0 = 1$\, MK  and $\rho_0 = 1.67 \times 10^{-12}$\, kg\, ${\text m}^{-3}$. Likewise, they found that in the considered heating and cooling model the polytropic index [$\gamma_{\rm eff}$] is not affected much by a change in $a$- and $b$-parameters. \citep[c.f. Figure 8b in][]{2021SoPh..296..105P}. Comparing Figure 8 of \citet{2021SoPh..296..105P} and panels c and d of Figure 6 in the present work we also note that the values of the polytropic index and phase difference vary in very close ranges in both cases. We suggest that this could be due to the fact that in both the considered loops we have very close values of thermal, viscous, and radiative ratios. In the case shown in Figure 6 we have $d=0.019$, $e=0.001$, and $r=0.173$ while in the case as shown in Figure 8 of \citet{2021SoPh..296..105P} they had $d=0.022$, $e=0.001$, and $r=0.144$. This hints at a good consistency between the mathematical analysis of the present work and that of \citet{2021SoPh..296..105P}.

Considering the loop parameters as measured by \citet{2015ApJ...811L..13W} with $T_0 = 9$\, MK , $n_0 = 2.6 \times 10^{9}\, {\rm cm}^{-3}$ ($\rho_0 = 5.2 \times 10^{-12}$\, kg \,${\text m}^{-3}$), and $L = 180$\, Mm, we calculate $d=0.075$, $r=0.044$ (Model I), $r=0.08$ (Model II). In the case of thermal conductivity  as the only damping mechanism we find that $\Delta \phi \approx 43^\circ$ and $\gamma_{\rm eff} \approx 1.546$. Including the radiative losses (Model I) with constant heating we find that $\Delta \phi \approx 44^\circ$ and $\gamma_{\rm eff} \approx 1.555$. Similarly for the radiative Model II with constant heating we calculate $\Delta \phi = 46.28^\circ$ and $\gamma_{\rm eff} = 1.55$. This suggests that radiative losses cannot explain the observed phase shift $\Delta \phi \approx 12^\circ$ and $\gamma_{\rm eff} \approx 1.64$ \citep{2015ApJ...811L..13W}. Further the consideration of a guessed heating function $a= -0.5, b=-3$ shows not much effect either as we calculate that in this case $\Delta \phi \approx 46.5^\circ$, $\gamma_{\rm eff} \approx 1.547$ (for radiative Model I) and $\Delta \phi = 51.25^\circ$, $\gamma_{\rm eff} = 1.53$ (for radiative Model II). This clearly shows that the observed values cannot be explained with the combined role of thermal conductivity and radiative cooling along with a temperature- and density-dependent heating function ($\propto \rho^a T^b$) and the conclusion of suppressed thermal conductivity and enhanced viscosity suggested by  \citet{2015ApJ...811L..13W} is not changed. The weak effects of heating--cooling misbalance on the phase difference and polytropic index in hot flaring loops also validate the seismological method proposed by \citet{2019ApJ...886....2W} in determining the transport coefficients based on the parametric study with a nonlinear 1D MHD model.

Taking into account the combined effect of major dissipative effects, the present article provides a detailed interpretation of observed phase shifts of standing slow-mode waves based on a linear MHD model. We also performed a parametric study on a large range of coronal loops by studying the individual role of all the effects involved. The present work is expected to be an important basis for the future theoretical and observational studies of standing, slow MHD waves in coronal loops. We also used the radiative model proposed by \citet{2008ApJ...682.1351K} for our analysis and found that although the overall nature of phase shifts is consistent for Model I and II, there is a significant deviation in the precise values of phase difference and polytropic index  for the two cases especially at longer loop lengths.

The presented linear MHD model is constrained by the one-dimensional and infinite-field approximations which limits the analysis to a less realistic approach as compared to the cylindrical flux-tube models with finite plasma-$\beta$ \citep[c.f.][]{2017ApJ...849...62N,2021A&A...646A.155D}. In addition, the transverse structuring in density and temperature of coronal loops may lead to nonlinear mode coupling and wave leakage  \citep[c.f.][]{2012ApJ...754..111O,2021arXiv211110696O}. Modelling of coronal loops in two or three dimensions allows us to include several properties that might influence the wave damping and phase shifts. The inclusion of heating--cooling misbalance in more realistic 2D slab or 3D loop models would be an important extension of the study of phase shifts presented here.

}
\begin{fundinginformation}
A.K. Srivastava acknowledges the support of UKIERI (Indo-UK) research grant and the ISSI-BJ regarding the science team project on ``Oscillatory Processes in Solar and Stellar Coronae". A. Prasad and K. Sangal acknowledge IIT-BHU for the support of computational facility in the present research.
The work of T.J. Wang was supported by NASA grants 80NSSC18K1131 and 80NSSC18K0668 as well as the NASA Cooperative Agreement 80NSSC21M0180 to CUA.  
\end{fundinginformation}

\begin{dataavailability}
The datasets generated during and/or analysed during the current study are available from the corresponding author on reasonable request.
\end{dataavailability}

\begin{ethics}
\begin{conflict}
The authors declare that they have no conflicts of interest.
\end{conflict}
\end{ethics}



\bibliographystyle{spr-mp-sola}
\bibliography{sola_bibliography_example}

\end{article} 

\end{document}